\documentclass[usenatbib,letter]{mnras}

\usepackage{newtxtext,newtxmath}

\usepackage[T1]{fontenc}
\usepackage{ae,aecompl}

\usepackage{graphicx}	
\usepackage{amsmath}	
\usepackage{amssymb}	
\usepackage{xspace}
\usepackage{xcolor}
\usepackage{aas_macros}
\usepackage{booktabs}
\usepackage[position=top]{subfig}
\usepackage{enumitem}
\usepackage[normalem]{ulem}

\usepackage{xpatch}

\usepackage{hyperref}
\usepackage{cleveref}

\makeatletter

\newcommand{\SFE}{\epsilon_{\mathrm{SF}}\xspace}
\newcommand{\SFEmin}{\epsilon_{\mathrm{min}}\xspace}
\newcommand{\Mcl}{M_{\mathrm{cl}}}

\newcommand{\Rcl}{R_{\mathrm{cl}}}

\newcommand{\Msun}{\mathrm{M}_{\odot}\xspace}

\newcommand*\diff{\mathop{}\!\mathrm{d}}

\newcommand{\Lw}{L_{\mathrm{mech}}}

\newcommand{\kms}{km\,s$^{-1}$\xspace}
\newcommand{\ccm}{cm$^{-3}$\xspace}
\newcommand{\rhocore}{\rho_{0}}
\newcommand{\rcore}{R_{0}}
\newcommand{\SFEff}{\epsilon_{\mathrm{ff}}}
\newcommand{\tff}{t_{\mathrm{ff}}}
\newcommand{\alp}{\tilde{\alpha}}

\defcitealias{Weaver1977}{W77}
\defcitealias{Rahner2017}{R17}

\title[WARPFIELD 2.0: Minimum star formation efficiencies]{WARPFIELD 2.0: Feedback-regulated minimum star formation efficiencies of giant molecular clouds}

\author[Rahner et al.]{Daniel 
Rahner$^{1}$\thanks{daniel.rahner@uni-heidelberg.de},  Eric W.\ 
Pellegrini$^{1}$, Simon C. O. Glover$^{1}$, Ralf S. Klessen$^{1,2}$ \\
$^{1}$ Universit{\"a}t Heidelberg, Zentrum f{\"u}r Astronomie, Institut f{\"u}r Theoretische Astrophysik, \\
Albert-Ueberle-Stra{\ss}e 2, 69120 Heidelberg, Germany\\
$^{2}$ Universit{\"a}t Heidelberg, Interdisziplin{\"a}res Zentrum f{\"u}r Wissenschaftliches Rechnen,\\
Im Neuenheimer Feld 205, 69120 Heidelberg, Germany}

\date{Accepted XXX Received YYY; in original form ZZZ}

\pubyear{2018}
\volume{XXX}

\hypersetup{draft}

\begin{document}

\label{firstpage}
\pagerange{\pageref{firstpage}--\pageref{lastpage}}
\maketitle

\begin{abstract}
Star formation is an inefficient process and in general only a small fraction of the gas in a giant molecular cloud (GMC) is turned into stars. This is partly due to the negative effect of stellar feedback from young massive star clusters.  In a recent paper, we introduced a novel 1D numerical treatment of the effects of stellar feedback from young massive clusters on their natal clouds, which we named {\sc warpfield}. Here, we present version 2 of the \textsc{warpfield} code, containing improved treatments of the thermal evolution of the gas and the fragmentation of the feedback-driven shell. As part of this update, we have produced new cooling and heating tables that account for the combined effects of photoionization and collisional ionization on the cooling rate of the gas, which we now make publically available.  
We employ our updated version of \textsc{warpfield} to investigate the impact of stellar feedback on GMCs with a broad range of masses and surface densities and a variety of density profiles. We show that the minimum star formation efficiency $\SFEmin$, i.e.\ the star formation efficiency above which the cloud is destroyed by feedback and further star formation is shut off, is mainly set by the average cloud surface density. A star formation efficiency of 1-6\,\% is generally sufficient to destroy a GMC. We also find star formation efficiencies per free-fall time $\SFEff \sim 0.3$\,\%, in good agreement with recent observations.  Our results imply that stellar feedback alone is sufficient to explain the low observed star formation efficiencies of GMCs. Finally, we show that very massive clouds with steep density profiles -- possible proxies of the giant clumps observed in galaxies at  $z \approx 2$ -- are more resilient to feedback than typical GMCs, with $\SFEmin$ between 1 and 12\,\%.
\end{abstract}


\section{Introduction}
One of the most important questions in the study of star formation in a galactic context is what determines the star formation activity of giant molecular clouds (GMCs). Both observations and theory agree that at least part of the answer is stellar feedback \citep[see e.g.][]{MacLow2004,McKee2007,Hennebelle2012,Krumholz2014,Molinari2014,Klessen2016}. When massive stars form in GMCs, they disrupt their birth environment via powerful stellar winds and the emission of ionizing radiation. After several million years, the same massive stars end their lives in supernova (SN) explosions. Often the combination of these stellar feedback processes is sufficient to overcome the gravitational attraction of the gas, and consequently the parental cloud is destroyed and further star formation inhibited \citep{Murray2010, Wang2010, Silich2013, Rahner2017}. If on the other hand stellar feedback is too weak, the remaining cloud material can re-collapse and form a new generation of stars \cite[e.g.][]{Wuensch2017,Szecsi2018,Rahner2018}.

A convenient way to quantify how effectively the GMCs in a given galaxy form stars is via the star formation efficiency $\SFE$. Although there are various different ways to define this quantity, the simplest, and the one which we use in this paper, is to define $\SFE$ simply as the fraction of the gas associated with a given GMC that is turned into stars during the lifetime of that GMC. Galactic observations yield values for $\SFE$ that are typically in the range of $1\,\% - 10\,\%$ for GMCs in the mass range  $10^5 - 10^7\,\Msun$ \citep[see e.g.][]{Murray2011,Vutisalchavakul2016}, while extragalactic observations yield values consistent with the lower end of this range if one assumes that a typical GMC survives for a few dynamical times \citep{Leroy2017,Kreckel2018}. 

An obvious question is how these observationally determined values for $\SFE$ compare with the minimal star formation efficiency necessary for feedback to disrupt the cloud, $\SFEmin$. Analytic models by \citet{Fall2010} and \citet{Kim2016} hinted at somewhat higher values for $\SFEmin$ than what is typically observed for $\SFE$, a result which taken at face value would suggest that feedback is {\em not} the process primarily responsible for destroying the clouds. These models, however, either did not include mechanical feedback (i.e.\ winds and SNe) or accounted for it only in a very simplistic manner, as their primary focus was radiative feedback. As shown in \citet{Rahner2017}, mechanical and radiative feedback are intrinsically linked and generally both need to be included to correctly model the dynamical evolution of massive clouds under the influence of stellar feedback. 

When a stellar wind is launched from a massive star it  creates a hot bubble of ionized gas. If there are several massive stars in a dense cluster, these bubbles quickly overlap to form one large bubble that envelopes the entire cluster (except in cases where star formation simultaneously takes place in a large volume of the cloud which is filled by very dense ($n \gtrsim 10^5$ cm$^{-3}$) gas; see \citealt{Silich2017}). As discussed by \citet{Castor1975} and \citet{Weaver1977}, the wind-blown bubble consists of several distinct parts. An inner low-density region, where the wind ejecta stream freely outwards, is surrounded by a region of shocked, hot gas ($10^5 - 10^8$\,K), which is bounded by an inner shock radius $R_1$ and an outer shock radius $R_2$. The material in this shocked region is mainly gas that has evaporated from the comparatively cold ($10^2 - 10^4$\,K) surrounding shell of swept-up interstellar medium. Beyond the shell lies material of the natal cloud that is still unaffected by the stellar feedback. The standard picture assumes that the surrounding gas is in hydrostatic equilibrium and follows a roughly constant density profile.  Leakage through holes and channels in the shell and radiative cooling of shocked material gradually leads to a transition from energy-driven expansion (where the thermal pressure of the bubble dominates) to momentum-driven expansion (where ram-pressure and radiation pressure dominate). In either case, pressure from winds (and later, SNe), i.e. thermal pressure of shock-heated gas at early times and ram-pressure at late times, is important for determining the coupling between the radiation and the gas and hence cannot be neglected. It governs the efficiency of radiation as a feedback source.

The optical depth of the gas inside the wind bubble is small and so the radiation from the central stellar cluster can easily reach the dense shell of swept-up material \citep{Townsley2003,Gupta2016}. Those photons with energies above $13.6\,$eV photoionize hydrogen in this shell. Depending on the density structure and chemical composition, either the entire shell becomes ionized or only the inner layers are affected \citep{Martinez-Gonzalez2014, Rahner2017}. This has strong consequences for the optical depth and the resulting photon escape fraction at different frequencies. The absorbed photons exert a pressure force on gas and dust and push the shell outwards. 

In \citet{Rahner2017}, we introduced a new 1D stellar feedback model, called {\sc warpfield}, which was designed to model the evolution of such a bubble and its impact on the surrounding ISM in a self-consistent fashion. As explained in more detail in that paper, this involves solving simultaneously for the dynamics of the bubble and surrounding dense shell and for the structure of the shell. This is necessary, as the dynamical state of the shell -- specifically, the pressure acting on its inner edge -- influences its structure, while its structure determines how well radiation couples to the gas, and hence how effective radiation pressure is at driving the expansion of the shell. This version of {\sc warpfield} has already been applied to model massive star forming regions such as 30 Doradus \citep{Rahner2018} and W49 \citep{Rugel2018}. However, it has the limitation that the transition from energy-driven expansion of the bubble to momentum-driven expansion is assumed to occur instantaneously, which is a significant simplification compared to the real physics of the problem.  
In this paper, we present version 2 of {\sc warpfield}, which removes this simplification. As an example of its use, we investigate how the value of $\SFEmin$ predicted by the code varies as  a function of cloud mass and surface density, and how these values compare with observationally determined values of $\SFE$. 

The paper is structured as follows.
First, in Section~\ref{sec:Dynamics}, we describe the main improvements that we have made to the model of \citet{Rahner2017}. 
One of the major improvements is the inclusion of a detailed model for the radiative cooling of the dense gas in the shell. This gas is strongly illuminated by the central stellar cluster, and its cooling therefore cannot be treated using standard ISM cooling curves, since these typically assume collisional ionization equilibrium. Instead, we have computed new cooling curves that account for the effects of both photoionization and collisional ionization that cover the parameter space relevant for cluster wind bubbles. These are presented in Section~\ref{sec:cooling}. 
Next, in Section~\ref{sec:Results}, we present several applications of the updated model. In Sections~\ref{sec:comparison} and \ref{sec:densprofiles}, we use the model to follow the evolution of feedback-driven shells which expand into GMCs with varying density profiles and compare to results from the previous version of \textsc{warpfield}. Finally, in Section~\ref{sec:minSFE} we present our results for the minimum star formation efficiencies of GMCs over a wide parameter range. The main results of this paper are summarized in Section~\ref{sec:conclusion}.

\section{Modelling the dynamics of a feedback-driven bubble} \label{sec:Dynamics}

Let us consider a GMC with a gas mass $M_{\mathrm{cl},0}$ that is turning some fraction of its gas into a massive star cluster in its centre with a mass of
\begin{equation}
M_* = \SFE M_{\mathrm{cl},0}.
\end{equation}
The remaining gas in the cloud, whose mass is now $M_{\mathrm{cl}} = M_{\mathrm{cl},0} - M_*$, is illuminated and accelerated by feedback from the central star cluster. Since we are only interested in massive clusters we use \textsc{starburst99} to model a cluster where the initial mass function (IMF) is fully sampled. We adopt a \citet{Kroupa2001} IMF with an upper stellar mass limit of 120\,M${_\odot}$. The individual stars follow Geneva evolutionary tracks for rotating stars \citep{Ekstrom2012, Georgy2012}. The relevant feedback properties of the star cluster are its total bolometric luminosity $L_{\mathrm{bol}}(t)$ and its mechanical luminosity
\begin{equation}
L_\mathrm{mech}(t) = \frac{1}{2}\dot{M}_* v_{\infty} ^2 .
\end{equation}
Here $\dot{M}_*$ is the mass loss rate of the cluster due to material being ejected by stellar winds or supernovae and $v_{\infty}$ is the terminal velocity of the ejecta.
The forces acting on the shell are due to the thermal pressure of the hot ionized inner bubble, the momentum input from radiation, and the ram-pressure from stellar winds and supernovae
\begin{equation}
F_\mathrm{ram}(t) = \dot{M}_* v_{\infty}.
\end{equation}
The opposing forces are gravity and ambient pressure which will slow down or even reverse the expansion of the shell. We consider GMCs in virial equilibrium and so we neglect ram pressure from inflow motions in the cloud for the time being.

\subsection{Expansion}
As feedback pushes the ISM away from the star cluster, a shell forms which consists of swept up gas.
The dynamics of the shell with radius $R_2$ and mass $M_{\mathrm{sh}}$ are given by the following set of ordinary differential equations (ODEs), the momentum and the energy equation,
\begin{eqnarray} \label{momentum_full}
\frac{\diff}{\diff t}\left(M_{\mathrm{sh}}\dot{R}_2\right) &=& 4\pi R_2^2(P_{\rm b}-P_{\mathrm{amb}}) - F_{\textup{grav}} + F_{\textup{rad}}, \\ \label{energy_w_cooling}
\dot{E}_{\rm b} &=& L_\mathrm{mech} - L_\mathrm{cool} - 4\pi R_2^2\dot{R}_2 P_{\rm b},
\end{eqnarray}
which we explain further below. (Note that here and elsewhere, dots denote differentiation with respect to time).

The pressure of the bubble $P_{\rm b}$ relates to its energy $E_{\rm b}$ via
\begin{equation} \label{pressure_correct}
P_{\mathrm{b}} = (\gamma -1)\frac{E_{\mathrm{b}}}{\frac{4\pi}{3}\left(R_2^3 - R_1^3 \right)}
\end{equation}
where $\gamma$ is the adiabatic index. We will use $\gamma = 5/3$ as appropriate for an ideal monatomic gas. The inner shock radius $R_1$ is set by a pressure equilibrium between the free-streaming winds and the hot bubble, from which immediately follows the implicit equation
\begin{equation} \label{R1}
R_1 = \left[ \frac{F_{\mathrm{ram}}}{2 E_{\rm b}}\left(R_2^3-R_1^3\right)\right]^{1/2} .
\end{equation} 
The ambient pressure $P_{\mathrm{amb}}$ outside of the shell is usually negligible in the regime we investigate here. However, if ionizing radiation from the star cluster escapes the confinement of the shell, it ionizes the ambient ISM, heating it to $T_{\rm i} \sim 10^4$\,K. At early times, when the immediate surrounding of the shell is very dense gas but ionizing radiation can still pass through the newly formed shell and ionize that gas, $P_{\mathrm{amb}}$ can be a large term. We thus use
\begin{equation}
   P_{\mathrm{amb}} =
   \begin{cases}
     P_0 + \frac{\mu_{\rm i}}{\mu_{\rm p}}n_{\mathrm{cl}}(R_2)kT_{\rm i} & \hspace{.5in} \text{if } f_{\rm{esc,i}} > 0, \\
     P_0  & \hspace{.5in} \text{otherwise}. \\
   \end{cases}
\end{equation}
Here, $\mu_{\rm i}$ and $\mu_{\rm p}$ are the mean molecular weights per ion and per particle, respectively ($\mu_{\rm i}/\mu_{\rm p} = 23/11$ for a composition with one helium atom per 10 hydrogen atoms), $n_{\mathrm{cl}}(R_2)$ is the number density of the cloud directly outside of the shell, and $f_{\rm{esc,i}}$ is the escape fraction of ionizing radiation through the shell.
The pressure $P_0$ of the ambient ISM in the absence of ionization depends on the galactic environment of the star-forming region.\footnote{The value of $P_0$ should be adjusted to a value appropriate for the studied environment. For the purpose of the results presented in Sec.~\ref{sec:Results}, we have taken $P_0$ to be negligible and set it to 0. However, this would not be an appropriate choice for studies of clusters in e.g.\ the Galactic Centre or a starburst galaxy.}

The forces of gravity and radiation pressure are given by
\begin{eqnarray} \label{grav-term}
F_{\rm{grav}} &=& \frac{GM_{\rm{sh}}}{R_2^2}\left( M_*+\frac{M_{\rm{sh}}}{2}\right), \\ \label{rad-term}
F_{\rm{rad}} &=& f_{\rm{abs}}\frac{L_{\rm{bol}}}{c} \left(1+\tau_{\rm{IR}}\right),
\end{eqnarray}
where $c$ is the speed of light. The fraction of radiation that is absorbed by the shell $f_{\rm{abs}}$ and its optical depth in the infrared $\tau_{\rm{IR}}$ are determined using a hydrostatic approximation for the shell (see \citealt{Rahner2017}).

In the previous version of \textsc{warpfield}, hereafter \textsc{warpfield1}, we neglected the energy lost to cooling until the age of the system reached the cooling time $t_{\mathrm{cool}}$ and assumed that all the energy was lost thereafter. The cooling time is defined as
\begin{equation} \label{cooltime}
t_{\rm{cool}} = 16\,\textup{Myr} \times  (Z/Z_{\odot})^{-35/22} 
n_{\rm{cl}}^{-8/11} L_{38}^{3/11},
\end{equation}
 with $Z$ being the metallicity, $n_{\rm{cl}}$ the cloud density in cm$^{-3}$, and $L_{38}=L_{\rm{mech}}/(10^{38}$\,erg\,s$^{-1}$) \citep{MacLow1988}. However, eq.~(\ref{cooltime}) only holds for constant density profiles, and even then ignoring cooling when $t < t_{\rm{cool}}$ and assuming immediate loss of all energy for $t \geq t_{\rm{cool}}$ is a major simplification \citep{Gupta2016}. Here, instead, we couple the energy loss term due to cooling $L_\mathrm{cool}$ to the energy equation. The loss term is given by
\begin{equation} \label{Lcool}
L_\mathrm{cool} = 4\pi \int\limits_{R_1}^{R_2} \frac{\diff U}{\diff t}\Big|_{\mathrm{rad}} r^2  \textup{d}r, 
\end{equation}
where $U$ is the internal energy density and where the integration runs from the inner shock at a radius $R_1$ to the outer radius of the bubble $R_2$, which is also the radius of the thin shell.
The rate of change of the radiative component of the internal energy density is given by
\begin{eqnarray} \nonumber
\frac{\diff U}{\diff t}\Big|_{\mathrm{rad}} &=& n n_\mathrm{e}\left[\Gamma\left(T, \ldots \right) - \Lambda\left(T, \ldots \right) \right] \\ \label{radenerg}
&=& - n n_\mathrm{e} \Lambda_{\mathrm{net}}\left(T, \ldots \right),
\end{eqnarray}
where $n$ and $n_\mathrm{e}$ are the ion and electron density, and $\Gamma$ and $\Lambda$ are the heating and cooling functions. We have used the dots to indicate that in general $\Gamma$ and $\Lambda$ are dependent on many parameters (and not just $T$) as we discuss in Sec. \ref{sec:cooling}.
Assuming that the pressure inside the bubble is independent of radius $r$ and neglecting magnetic fields, we have
\begin{equation}
n(t, r) = \frac{\mu_{\rm p}}{\mu_{\rm i}}\frac{P_{\rm b}(t)}{kT(t, r)},
\end{equation}
where again $r$ runs from $R_1$ to $R_2$.

\citet{Weaver1977} describes a procedure to calculate the temperature profile inside the bubble when thermal conduction between bubble and shell is taken into account. In short, $T$ is a function (which is described in detail in Appendix \ref{sec:Bubble_Struc}) of $t, r, R_2, \dot{R}_2, E_{\mathrm{b}}, \dot{E}_{\mathrm{b}}$, and $\dot{T}_\xi$,
where $T_\xi$ is the temperature measured at some fixed scaled radius inside the bubble, e.g. $\xi \equiv r/R_2 = 0.9$, i.e. 
\begin{equation} \label{Txi}
T_\xi = T(t, r=\xi R_2, R_2,\dot{R}_2, E_{\mathrm{b}}, \dot{E}_{\rm b}, \dot{T}_\xi) .
\end{equation}

It becomes clear that inserting $T$ into Eqs.~(\ref{Lcool}) and (\ref{energy_w_cooling}) leads to $\dot{E}_{\rm b}$ being given only implicitly. Furthermore, $\dot{T}_\xi$ is also only given implicitly by Eq.~(\ref{Txi}), that is, we have $\dot{E}_{\rm b} = \dot{E}_{\rm b}(\ldots, \dot{E}_{\rm b}, \dot{T}_\xi)$ and $T_{\xi} = T_{\xi}(\ldots, \dot{E}_{\rm b}, \dot{T}_\xi)$.
In order to solve for the dynamical evolution of the shell it is thus necessary not only to augment the ODE system by Eq.~(\ref{Txi}), but in addition -- due to the implicit nature of Eqs.~(\ref{energy_w_cooling}) and (\ref{Txi}) -- for each time step a root finding algorithm must be used to find the values of $\dot{E}_{\rm b}$ and $\dot{T}_\xi$ so that Eqs.~(\ref{energy_w_cooling}) and (\ref{Txi}) are simultaneously satisfied. Such a scheme is implemented in the version of \textsc{warpfield} presented here.

The system of ODEs which fully describes the dynamics of bubble and shell, Eqs.~(\ref{momentum_full}), (\ref{energy_w_cooling}), and (\ref{Txi}), is stiff. In order to solve it, \textsc{warpfield} makes use of the \textsc{scipy} routine \textsc{solve\_ivp} which wraps around the \textsc{fortran} solver \textsc{lsoda} \citep{Hindmarsh1983, Petzold1983}.

\subsection{Stalling}

\subsubsection{Shell Fragmentation} \label{sec:frag}
Cooling is not the only way for the bubble to lose energy. As soon as the shell surrounding the bubble fragments, the hot gas can leak out, introducing a second loss term. This phase of leakage is hard to describe analytically \citep[cf.][]{Harper-Clark2009}. The speed with which the energy leaks out is set by both the pressure inside the bubble and the size of the holes in the shell, which both vary as function of time. Furthermore the assumption of pressure equilibrium used for the calculation of $T(r)$ becomes invalid and the previously outlined method to calculate $L_{\mathrm{cool}}$ breaks down.

Here we employ a simplified treatment of leakage 
but note that this a weakness of the model which we plan to address in future work. In order to prevent leakage when the bubble is still deeply embedded in the cloud, we allow this process to occur only when the shell radius has reached 10\,\% of the cloud radius, i.e.
\begin{equation}
R_2 \geq 0.1R_{\mathrm{cl}} .
\end{equation}
In our model, fragmentation occurs when the above, necessary condition as well as one of the following additional conditions is fulfilled:
\begin{enumerate}
\item \textbf{Gravitational fragmentation} occurs when inside a region cut from the surface of the shell the combined kinetic divergent energy due to stretching as it expands and the thermal energy are outweighed by the gravitational binding energy of that region \citep{McCray1987, Ostriker1981}. This is the case when
\begin{equation}
0.67 \frac{3GM_{\mathrm{sh}}}{4\pi\dot{R}_2R_2 c_\mathrm{s,sh}}  > 1 .
\end{equation}
Here $c_\mathrm{s,sh}$ is the lowest sound speed of the shell (typically $1\,$\kms, unless the shell is fully ionized). \\
\item \textbf{Rayleigh-Taylor (RT) instabilities} occur when a dense fluid is accelerated with respect to a fluid of lower density. As the shell decelerates while it sweeps up the cloud, radiation pressure counteracts the formation of RT instabilities. However, when the strength of feedback increases during the Wolf-Rayet phase and after the first SN explosions occur, the shell can accelerate again (if the density of the surrounding material is low enough). So, we take
\begin{equation}
\ddot{R}_2 > 0 , 
\end{equation}
as the criterion for the occurrence of RT instabilities and shell fragmentation. \\
\item \textbf{Density inhomogeneities} are hard to model in a 1D code. As in \citet{Rahner2017} we assume that the lower density parts of the shell are opened up to feedback channels as soon as the parental cloud is swept up completely, i.e.
\begin{equation}
R_2 > R_{\mathrm{cl}} .
\end{equation} 
However, we note that this will depend on the structural details of the cloud which is being modelled.
\end{enumerate}
The fragmentation of the shell via one of these processes at a time $t_{\mathrm{frag}}$ marks the end of the energy-driven expansion. Starting at $t = t_{\mathrm{frag}}$, we remove the energy from the bubble over a sound crossing time
\begin{equation} \label{Edot_transition}
\dot{E}_{\rm b} = - \frac{E_{\rm b}(t_{\mathrm{frag}})}{t_{\mathrm{s,cr}}(t_{\mathrm{frag}})}
\end{equation}
with
\begin{equation}
t_{\mathrm{s,cr}} = \frac{R_2}{c_{\mathrm{s, b}}}
\end{equation}
where $c_{\mathrm{s, b}}$ is the sound speed corresponding to the volume-averaged temperature of the bubble. Usually, $t_{\mathrm{frag}}$ is of the order $0.1 - 1$\,Myr for GMCs investigated here (see Sec. \ref{sec:minSFE}).

During this phase of energy leakage, Eq.~(\ref{Edot_transition}) replaces Eq.~(\ref{energy_w_cooling}).
It marks the transition between the energy-driven and momentum-driven phases and lasts until all energy has been removed from the bubble, i.e.\ $E = 0$, which is the case at $t=t_{\mathrm{frag}} + t_{\mathrm{s,cr}}(t_{\mathrm{frag}})$. Afterwards, ram-pressure from winds and SNe hits the shell without an intervening layer of shocked gas. From then on, the dynamical evolution of the shell is controlled solely by the momentum equation with $4\pi R_2^2 P_{\rm b} = F_{\mathrm{ram}}$. 

\subsection{Collapse and sequential star formation} \label{sec:collapse}
Should at any point in time the inward directed terms in Eq.~(\ref{momentum_full}) outweigh the outward directed terms, the shell loses momentum. In particular, this can happen as the bubble's energy is lost via cooling or leakage and its pressure drops dramatically. The pressure also drops soon after the death of the most massive stars when even the pressure from SN explosions of the slightly less massive stars is not sufficient to replace the missing stellar wind feedback. This is the case when the cluster is approximately $4-5$\,Myr old. If the shell starts to collapse back onto the central star cluster, we keep the shell mass constant and follow the evolution until the shell radius has shrunk to the ``collapse radius'' $R_{\mathrm{coll}} = 1$\,pc. 
Collisions between fragmented clumps of the shell will induce the birth of new stars and so we take the time of re-collapse to mark the time $t_\mathrm{SF}$ of a new star formation event. Sequential star formation was already implemented in \textsc{warpfield1} and has been used to model multiple stellar populations in 30~Dor \citep{Rahner2018}. In the more general treatment present in \textsc{warpfield2}, a new star cluster forms according to one of the following two prescriptions:
\begin{enumerate}
\item The new star cluster forms with the same star formation efficiency $\SFE$ as the first cluster. Since the cloud is less massive than it was before the previous cluster formed the new cluster will be less massive as well:
\begin{equation}
M_{*,i} = M_{*,i-1}(1-\SFE),
\end{equation}
where $M_{*,i}$ denotes the cluster mass of the $i$-th generation. \\
\item We form the next star cluster with a fixed star formation efficiency per free-fall time $\SFEff$ \citep[see][]{Krumholz2005,Krumholz2007}, where we use the free-fall time $\tff$ which corresponds to the mean density of the cloud $\bar{\rho}$. The first cluster forms with $\epsilon_{\mathrm{SF},1} = \SFEff$. Each subsequent star formation efficiency is given by
\begin{equation}
\epsilon_{\mathrm{SF},i} =  \frac{t_{\mathrm{SF}, i} - t_{\mathrm{SF}, i-1}}{\tff} \SFEff.
\end{equation}
This also allows the later clusters to form with higher masses than the previous cluster if the time difference between two star formation events is larger than the free-fall time of the cloud.
\end{enumerate}
In either case, at the time a new star cluster forms, we reset the cloud structure and distribute the ISM according to the same density profile as before.
The gas is now subject to feedback $\mathcal{F} \equiv \{L_{\mathrm{bol}}, \Lw, F_{\mathrm{ram}} \}$ originating from several generations of stars
\begin{equation}
\mathcal{F}(t) = \sum\limits_{i=1}^{N_{\mathrm{gen}}(t)}\mathcal{F}_i(M_{*,i},t-t_{\mathrm{SF}, i}),
\end{equation}
where $N_{\rm{gen}}(t)$ is the number of cluster generations present at time $t$. Again, the expansion is initially driven mostly by energy and later, after the hot gas has cooled and leaked out of the bubble, by momentum. Should feedback again be insufficient to overcome gravity, another cluster forms and so on, until the shell dissolves.
We regard the shell as dissolved when its maximum density has dropped below 1\,\ccm for a duration of at least 1\,Myr or when it has expanded to 1\,kpc.\footnote{At this point, we assume galactic shear to have disrupted the cloud.} The age of the cloud when it dissolves marks its lifetime $t_{\mathrm{life}}$. We also stop the simulation when 30\,Myr have passed, which corresponds to lifetime estimates of GMCs in the Large Magellanic Cloud \citep[LMC,][]{Kawamura2009}.

\section{Cooling} \label{sec:cooling}
The wind bubble is of double importance for the evolution of the shell. 
The over-pressured bubble not only pushes the ISM outwards, but also sets the density at the inner shell boundary, thus determining the coupling between radiation and the shell (see Eq. \ref{rad-term} and \citealt{Rahner2017} for more details). How the pressure of the bubble changes as a function of time depends on the amount of energy lost via radiative cooling. 
This in turn is dependent on the temperature profile of the bubble and on the net cooling function $\Lambda_{\mathrm{net}}$ (see Eqs. \ref{Lcool} and \ref{radenerg}). 
Therefore, an accurate treatment of cooling is of paramount importance to solving the evolution of the HII region. 

Given a known set of elemental abundances in the gas -- which in our case are fixed once we specify the metallicity -- the value of $\Lambda_{\mathrm{net}}$ depends primarily on two things: the temperature of the gas and its overall ionization balance (i.e.\ for each element, what fraction is neutral, what fraction is singly ionized, etc.). In the simple case of collisional ionization equilibrium (CIE), the ionization balance itself depends solely on the temperature\footnote{There is no density dependence because both collisional ionization and radiative recombination have $n^{2}$ dependencies on density, and so changes in the density affect both equally, leaving the ionization balance unaffected.}, and so $\Lambda_{\mathrm{net}}$ to a good approximation depends only on $T$. However, in our case, the bubble sits in close proximity to the star cluster, whose radiation is not shielded by the intervening ISM. Consequently, the influence of ionizing radiation from the star cluster is substantial and the gas is often not in CIE. In this case, the ionization balance, and hence the cooling rate, is determined by four parameters: the gas temperature $T$, the gas density $n$, the flux of ionizing photons $\Phi_{\rm i}$, and the age of the stellar cluster $t_{\rm age}$. The temperature controls the collisional ionization and radiative recombination rate coefficients, as before, while the density and ionizing photon flux determine the relative importance of radiative recombination and photoionization. The age of the stellar cluster is important as it determines the spectral shape of the incident radiation, and this in turn plays a major role in controlling the impact that it has on the ionization balance of the gas. For example, if we are interested in the photoionization of hydrogen, which requires photons with energies $\geq 13.6$~eV, or of metals with ionization potentials below that of hydrogen, then radiation from a large population of older B stars can be competitive with that from a small number of younger, brighter O stars. On the other hand, the production of O$^{2+}$, which has an ionization potential of 35~eV, requires photons from stars with effective temperatures above 36000~K and hence is dominated by emission from O stars.

To deal with this complexity, we use \textsc{cloudy} \citep{Ferland2017} to estimate $\Lambda_{\mathrm{net}}$ for a variety of different values of the main controlling parameters. In the simulations presented here, we consider two different metallicities, ${\rm Z = 0.014 = Z_{\odot}}$ and ${\rm Z = 0.002 = 1/7 \, Z_{\odot}}$, and a range of values for the gas temperature $T$, the number density $n$, the incident flux of ionizing photons $\Phi_{\rm i}$, and the age of the stellar cluster $t_{\rm age}$. 
The range of values considered for each parameter is summarized in Table~\ref{tab:cooling_grid}. Further details of the \textsc{cloudy} models are provided in Appendix~\ref{sec:cooling_appendix}. We have made the cooling curves themselves publicly available, and information on how to access them is also given in Appendix~\ref{sec:cooling_appendix}.

We note that the basic idea behind this approach is not new. For example, a similar, but much larger grid of cooling curves has been published by \citet{Gnedin2012}. However, in that work the stellar component of the considered spectral energy distributions was chosen to represent the interstellar radiation field of the Milky Way, i.e.\ mainly old stars that evolved with a continuous 
star formation history over the course of 1\,Gyr. This is very different to what is necessary for our purpose where we need the spectrum of a young star cluster in which all stars are coeval or are born in only a few distinct populations separated by no more than a few tens of Myr, such as Sandage-96, 30~Doradus, and the Orion Nebula Cluster \citep{Vinko2009,Sabbi2012,Beccari2017}. Similarly, the tabulated cooling curves bundled with the {\sc grackle} chemistry code \citep{Smith2017} or computed by \citet{Wiersma2009} or
\citet{Emerick2018} account for both photoionization and collisional ionization, but consider radiation fields designed to represent the extragalactic background, rather than the spectrum of a young massive cluster, making them unsuitable for our purposes. 

The parameter space covered by our models has been tailored to our regime of interest, namely shells driven by hot bubbles inside dense GMCs. There is thus no need to calculate the cooling function for hot gas illuminated by a cluster older than 10\,Myr, as the HII region around such a cluster will no longer be expanding in the energy-driven limit. The temperature of the ISM which is directly illuminated by such a cluster will be constant at $\sim 10^4$\,K (see Sec. \ref{sec:frag}). We also neglect shielding of radiation inside the bubble (but not inside the shell). This is well justified because the column density of the bubble is low. For the stars we use Pauldrach/Hillier atmospheres, i.e.\ \textsc{wm-basic} models \citep{Pauldrach2001} when the star cluster is younger than 3\,Myr and \textsc{cmfgen} models \citep{Hillier1998} thereafter. The elemental composition of the ISM has been chosen to represent that observed in HII regions \citep[see Table 7.2 in][]{Ferland2013b}.
At $Z=0.002$ we use the same ISM composition as at $Z=0.014$ (solar) but scale all abundances down by a factor 1/7.

\begin{table}
\begin{tabular}{|l|c|c|c|c|}
\hline 
  & min & max & step size & remark\\ 
\hline 
$\log T$/K & 3.5 & 5.5 & 0.1 & higher $T$: CIE\\

$\log n$\,/cm$^{-3}$ & -4 & 12 & 0.5 & \\ 

$\log \Phi_\mathrm{i,cgs}$ & 0 & 21 & 1 & with cosmic rays\\ 

$t_{\mathrm{age}}$/Myr & 1 & 5 & 1 & also $t_{\mathrm{age}} = 10$\,Myr\\ 

$Z/Z_{\odot}$ & 1/7 & 1 & 6/7 & HII region abundances

\end{tabular} 
\caption{Overview of cooling grid. For each model the net cooling function $\Lambda_{\mathrm{net}}$ has been determined with \textsc{cloudy}. $T$: temperature, $n$: number density, $\Phi_{\mathrm{i,cgs}}$: number flux of ionizing photons in cm$^{-2}$\,s$^{-1}$, $t_{\mathrm{age}}$: age of the star cluster, $Z$: metallicity (we assume $Z_{\odot} = 0.014$).}
\label{tab:cooling_grid}
\end{table}

For the dynamical evolution of the system we use the presented grid as a lookup table and linearly interpolate $\Lambda_{\mathrm{net}}$ between grid points. For $T>10^{5.5}$\,K, CIE is a reasonable assumption and in this temperature range we employ tabulated cooling curves by \citet{Gnat2012} for $Z=0.014$ and \citet{Sutherland1993} for $Z=0.002$.
 In the case of multiple generations of star clusters (see Sec.~\ref{sec:collapse}), we use the spectral shape of the youngest cluster but scale it by the total flux of ionizing photons. For rotating stars this is a reasonable approximation as the spectral shape of the ionizing part of the spectrum of an older cluster only significantly differs from that of a very young cluster when the emission rate of ionizing photons has dropped by more than 1 order of magnitude.\footnote{We note that this does not hold true for non-rotating stars.}

\begin{figure}
\centering
\includegraphics[width=0.49\textwidth]{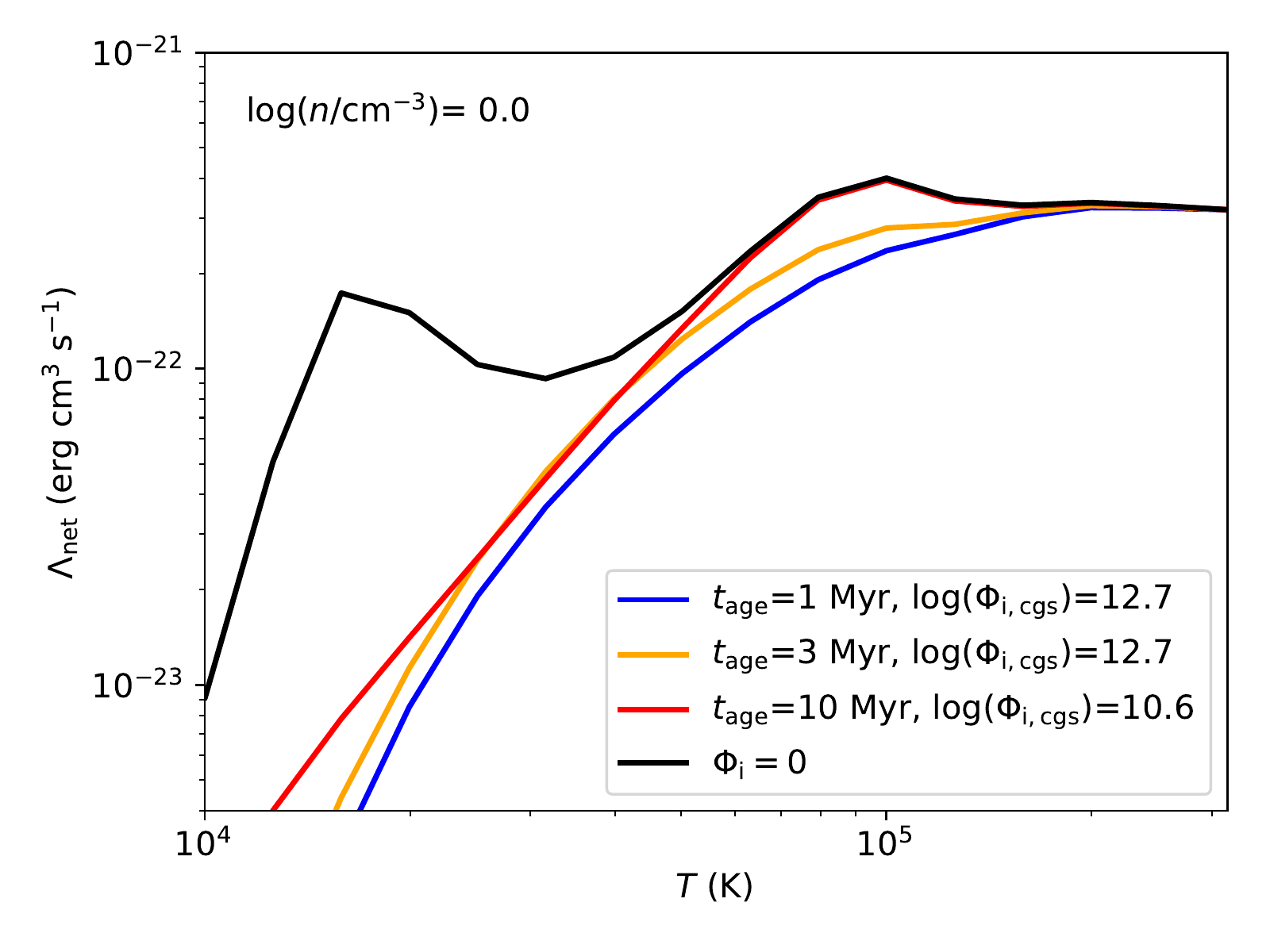}
\caption{Net cooling curves for solar metallicity (with elemental composition appropriate for HII regions) and a particle density of 1\,cm$^{-3}$. Cooling curves are shown by different colours for different ages $t_{\mathrm{age}}$ of the illuminating star cluster ($M_* = 10^6$\,$M_{\odot}$). The ionizing photon flux ($\Phi_\mathrm{i,cgs} = \Phi_\mathrm{i}$/(cm$^{-2}$\,s$^{-1}$))  has been calculated according to the time evolution of the cluster and at a distance of 10\,pc.}
\label{fig:CoolZ1}
\end{figure}

An example of non-CIE cooling curves at a density of $n=1\,$cm$^{-3}$ is shown in Fig. \ref{fig:CoolZ1}. At temperatures between $10^{4}$~K and $4 \times 10^{4}$~K, we see a large difference between the pure CIE cooling curve (black line) and the cooling curves for models with large ionizing fluxes, even for a relatively old stellar cluster. This difference is driven largely by the behaviour of the hydrogen, which dominates the CIE cooling rate at low temperatures. In the models with non-zero $\Phi_{\rm i}$, the hydrogen is far more ionized at low $T$ than in the pure CIE run, and consequently the contribution made by Lyman-$\alpha$ cooling to the total cooling rate is much smaller \citep[c.f.][]{Efstathiou1992}, with the bulk of the cooling now coming from dust which has evaporated from the shell into the bubble, bremsstrahlung, and various emission lines of oxygen and carbon. A similar but less pronounced effect is visible close to $10^{5}$~K, driven by changes in the He$^{+}$ abundance in the models with low $t_{\rm age}$. Notably, in the model with $t_{\rm age} = 10 \: {\rm Myr}$, the cooling curve at this temperature is the same as in the CIE case, as a cluster of this age produces very few photons capable of ionizing He$^{+}$ to He$^{2+}$. It is also apparent that above a few times $10^{5}$~K, the ionizing flux makes no difference to the cooling rate, since at these temperatures collisional ionization produces a more highly ionized state than can be produced by photoionization by stars. Further examples of cooling curves drawn from our set of models can be found in Appendix~\ref{sec:cooling_appendix} and in the online data.

We do not treat in detail heating and cooling in the feedback-driven shell but instead assume that the temperatures of the ionized and neutral phase are $10^4$ and $10^2$\,K, respectively. For the purpose of determining the approximate coupling between radiation and the ISM in the shell, which is necessary to determine the effect of radiation pressure, this is sufficient (for details see \citealt{Rahner2017}). For the modelling of emission lines, however, a detailed treatment of the chemistry inside the shell is indispensable. This can be carried out in a post-processing step, as we explore in detail elsewhere (Pellegrini et~al., in prep.)

\section{Results} \label{sec:Results}

\subsection{Comparison to WARPFIELD1} \label{sec:comparison}

With respect to the previously published version of \textsc{warpfield} \citep{Rahner2017}, hereafter referred to as \textsc{warpfield1}, several important improvements have been made, as discussed in the previous sections. In short, cooling and energy leakage of the hot bubble are now treated in a less simplified manner: In \textsc{warpfield1} cooling was treated as removing all energy at once when $t = t_{\mathrm{cool}}$ and fragmentation of the shell was only considered when $R_2 = R_{\mathrm{cl}}$. 
Even though we now allow the shell to fragment earlier due to RT or gravitational instabilities, radiative cooling is less efficient now in decreasing the thermal pressure of the bubble. While a large percentage of the energy is still radiated away, the retained energy is still sufficient to drive the expansion significantly more than pure momentum-driving would. Also, while the high pressure of the wind bubble is retained, the density of the shell remains high \citep[cf.][]{Rahner2017} and more ionizing radiation is absorbed by the shell, increasing the effect of radiation pressure as a source of feedback. The net result of these improvements is that in general stellar feedback is somewhat more efficient in pushing the gas outward and destroying the cloud. 

\begin{figure}
\centering
\includegraphics[width=1.0\linewidth]{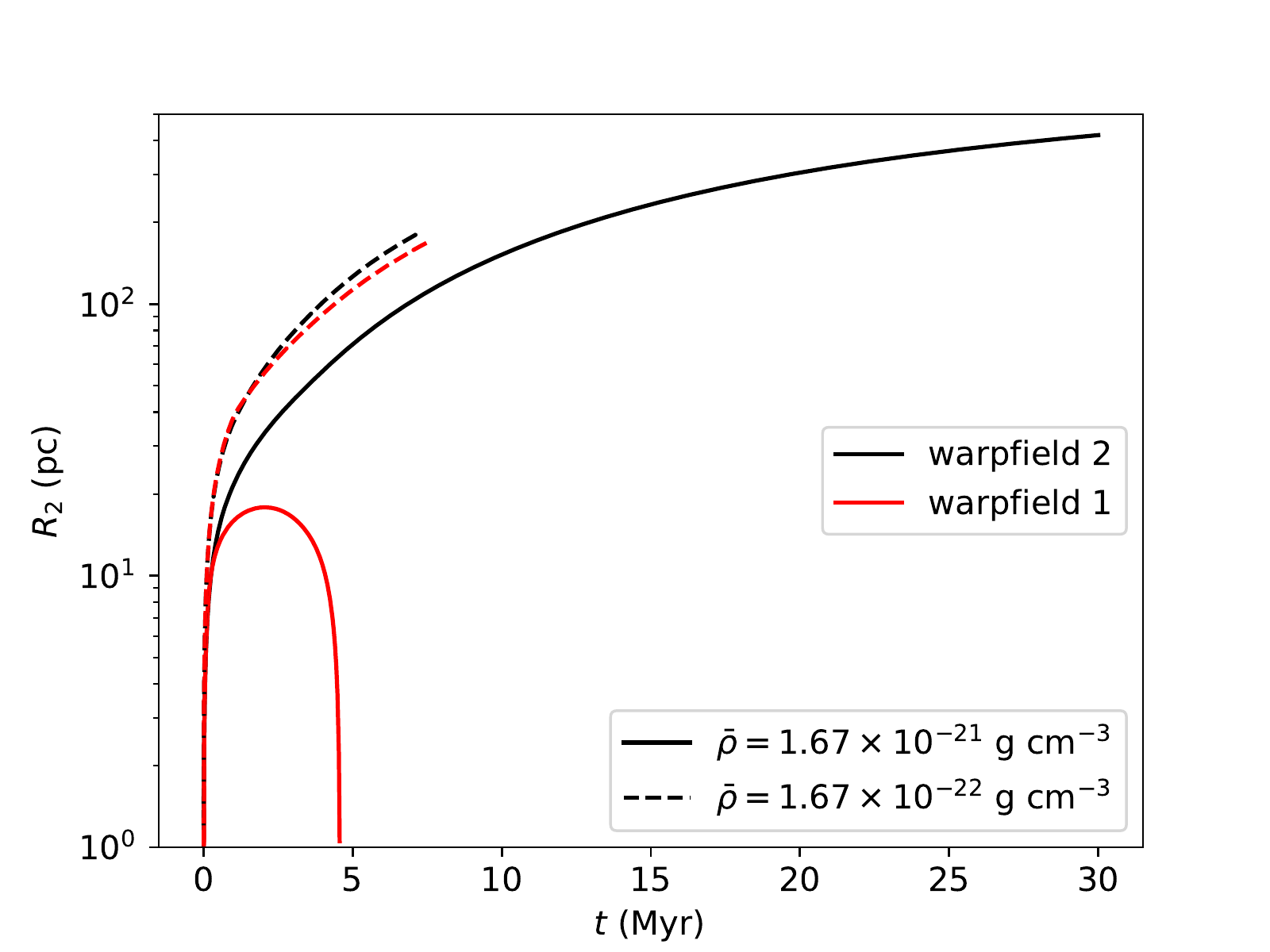}
\caption{Comparison between the evolution of the shell radius $R_2$ for models with $M_{\mathrm{cl}, 0} = 10^6$\,M$_{\odot}$, $\SFE = 0.05$, and two different mean densities, simulated with \textsc{warpfield1} \citep{Rahner2017} and \textsc{warpfield2} (this paper). The clouds have a constant density profile ($\alpha = 0$). }
\label{fig:exp_compold}
\end{figure}

Two examples where this can be seen are presented in Fig.~\ref{fig:exp_compold}. In cases where $\SFE$ is close to the minimum star formation efficiency $\SFEmin$ -- which we here define as the lowest star formation efficiency that is needed to destroy the cloud via stellar feedback after a \textit{single} star formation event -- the ``new'', somewhat stronger feedback can make the difference between continued expansion of the feedback-driven shell and a re-collapse of the ISM onto the star cluster. In cases where $\SFE$ is well above $\SFEmin$, the effect is a somewhat faster expansion and destruction of the cloud. Typically, the minimum star formation efficiencies we obtain with \textsc{warpfield2} are a few per cent lower than those presented in \citet{Rahner2017}, as will be shown in Section~\ref{sec:minSFE}.

\subsection{Variations in cloud density profile} \label{sec:densprofiles}

The other major difference compared to \textsc{warpfield1} is our treatment of the density distribution of the cloud. In \textsc{warpfield1}, this was assumed to be uniform, necessitated by our use of an analytic expression \citep{Bisnovatyi-Kogan1995} for the early, energy-driven phase, while in \textsc{warpfield2} we can treat any spherically symmetric density profile. In the following analysis, we consider not only homogeneous GMCs but also GMCs where the density $\rho_{\mathrm{cl}}$ follows a a power law profile with a homogeneous inner core:
\begin{equation}
   \rho_{\mathrm{cl}}(R) =
   \begin{cases}
     \rhocore & \hspace{.5in} \text{if } R \leq \rcore \\
     \rhocore \left( \frac{R}{\rcore}\right)^{\alpha}  & \hspace{.5in} \text{if } \rcore < R \leq R_{\mathrm{cl}}\\
     \rho_{\mathrm{amb}} & \hspace{.5in} \text{if } R > R_{\mathrm{cl}}.
   \end{cases}
\end{equation}

We limit ourselves to the range $-2 \leq \alpha \leq 0$, where the case $\alpha = 0$ corresponds to homogeneous clouds, while the steep density profile $\alpha = -2$ corresponds to so-called singular isothermal spheres. Such systems are interesting, because they are commonly encountered in the study of isothermal, self-gravitating systems \citep{Larson1969,Penston1969,Shu1977,Whitworth1985}. Clumps and cores forming at the stagnation points of large-scale convergent flows in the turbulent galactic ISM \cite[see, e.g.,][and references therein]{McKee2007,Klessen2016} typically have density profiles that mimic Bonnor-Ebert spheres \citep{Ebert1955,Bonnor1956} with a flat inner core and a smooth transition to an approximate $R^{-2}$ radial density profile at larger radii \citep{BP2003, Klessen2005}, and hence for our purposes may be well approximated by an $\alpha = -2$ profile. This is certainly the preferred density structure for low-mass prestellar cores that will eventually form individual stars  \cite[e.g.][]{Bacmann2000, Alves2001,Koenyves2010} and also seems applicable to high-mass systems \citep{Motte2018}. Once the central cluster has formed, the density profile is best fit by a power law with slope $\alpha = -1.5$ \cite[e.g.][]{Ogino1999} in the infalling envelope. Observations of molecular cloud clumps with embedded star clusters indicate exponents in the range  $-2 \leq \alpha \leq -1$ \citep[e.g.][]{Beuther2002}. For even higher masses, numerical simulations of star-forming giant clumps at redshift $z \approx 2$ also report values of $\alpha \approx -2$ \citep{Ceverino2012}. 
Because of these large variations, and also to account for the fact that the presence of turbulence can lead to significant deviations from simple analytic models and that the average density profile of the GMC as a whole may differ from that of the dense cores and clumps within it, we investigate a range of possible slopes and study the impact of this parameter on the dynamical evolution of the system and on the resulting minimum star formation efficiency, $\SFEmin$.  In the suite of models presented here,  we will concentrate on power law exponents $\alpha = 0, -1, -1.5, -2$ and consider a range of average cloud densities $\bar{\rho}$.

We set the core density to $\rhocore = 1.67\times 10^{-19}$\,g\,cm$^{-3}$, except in the case of $\alpha = 0$, where $\rhocore = \bar{\rho}$. The value for the core radius $\rcore$ follows from the condition that
\begin{equation}
M_{\mathrm{cl}} = 4\pi \int\limits_0^{R_{\mathrm{cl}}} R^2 \rho_{\mathrm{cl}}(R) \diff R .
\end{equation}
while the cloud radius $R_{\mathrm{cl}}$ is set by the cloud mass and the average density.
The density of the ambient ISM $\rho_{\mathrm{amb}}$ (i.e. the ISM beyond the cloud radius) is set to $1.67\times 10^{-25}$ \,g\,cm$^{-3}$ but changing this value by an order of magnitude has little impact on the eventual fate of the GMC \citep{Rahner2017}. Each model is thus uniquely specified by $M_{\mathrm{cl},0}, \SFE, \bar{\rho}, \alpha$, and $Z$ (although here we only consider $Z=Z_{\odot}$).

\begin{figure}
\centering
\includegraphics[width=1.0\linewidth]{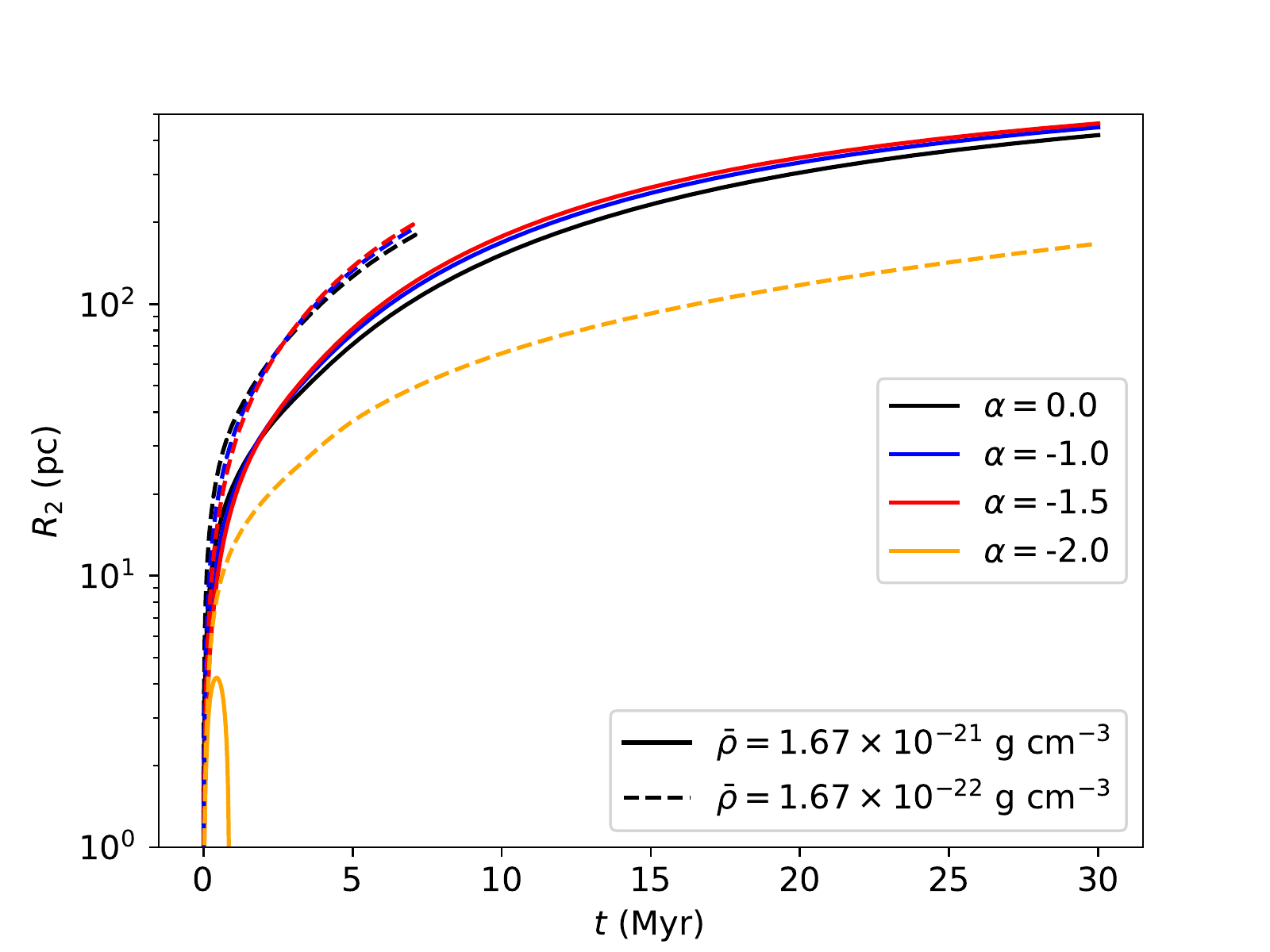}
\caption{Evolution of the shell radius for different different density slopes (colours) and initial average densities (line style). The models have $M_{\mathrm{cl}, 0} = 10^6$\,M$_{\odot}$, $\SFE = 0.05$.}
\label{fig:exp}
\end{figure}

In Fig. \ref{fig:exp}, we present the evolution of clouds with $\bar{\rho} = 1.67 \times 10^{-21} \, {\rm g \, cm^{-3}}$ and $\bar{\rho} = 1.67 \times 10^{-22} \, {\rm g \, cm^{-3}}$ and different density slopes. Whereas the behaviour of clouds with $-1.5 \leq \alpha \leq 0$ is very similar, clouds with even steeper density profiles are considerably harder to destroy with stellar feedback. 
This is partly due to the more negative gravitational binding energy of dense clouds which is defined as
\begin{equation}
E_{\mathrm{bind}} \equiv -G\int\limits_{M_{\mathrm{cl},0}} \frac{M(r)}{r} \diff m.
\end{equation}
For the density profiles investigated here ($\alpha \geq -2$), this becomes 
\begin{equation} \label{Ebind}
E_{\mathrm{bind}} = - \left[ U_0 + \frac{4\pi\rhocore G}{\rcore^{\alpha}} (u_1 + u_2) \right],
\end{equation}
with 
\begin{eqnarray}
U_0 &=& 2\pi \rhocore G M_* \rcore ^2 + \frac{16\pi^2}{15}\rhocore^2 G\rcore ^5, \\
u_1 &=& \left(M_0 - \frac{4\pi \rhocore}{3+\alpha}\rcore ^3\right) \times
\begin{cases}
\frac{\Rcl^{2+\alpha} - \rcore^{2+\alpha}}{2+\alpha}, \textup{\ if } \alpha \neq -2\\
\ln \left( \frac{\Rcl}{\rcore}\right), \textup{\ if } \alpha = -2,
\end{cases} \\
u_2 &=& \frac{4\pi \rhocore}{\rcore ^\alpha \left( 3 + \alpha \right)} \frac{\Rcl^{5+2\alpha} - \rcore^{5+2\alpha}}{5+2\alpha} .
\end{eqnarray}
Here, $M_0$ is the sum of the masses of the core and the star cluster.
Besides this obvious effect of making the material harder to unbind, the amount of cooling is also affected. As the bubble expands more slowly in a dense cloud, the density of the material accumulating inside the bubble is higher, leading to stronger radiative cooling and thus a further decrease in the efficiency of energy-driven mechanical feedback. 

\subsection{Minimum star formation efficiencies} \label{sec:minSFE}

\begin{figure*}
\centering
\includegraphics[width=0.48\linewidth]{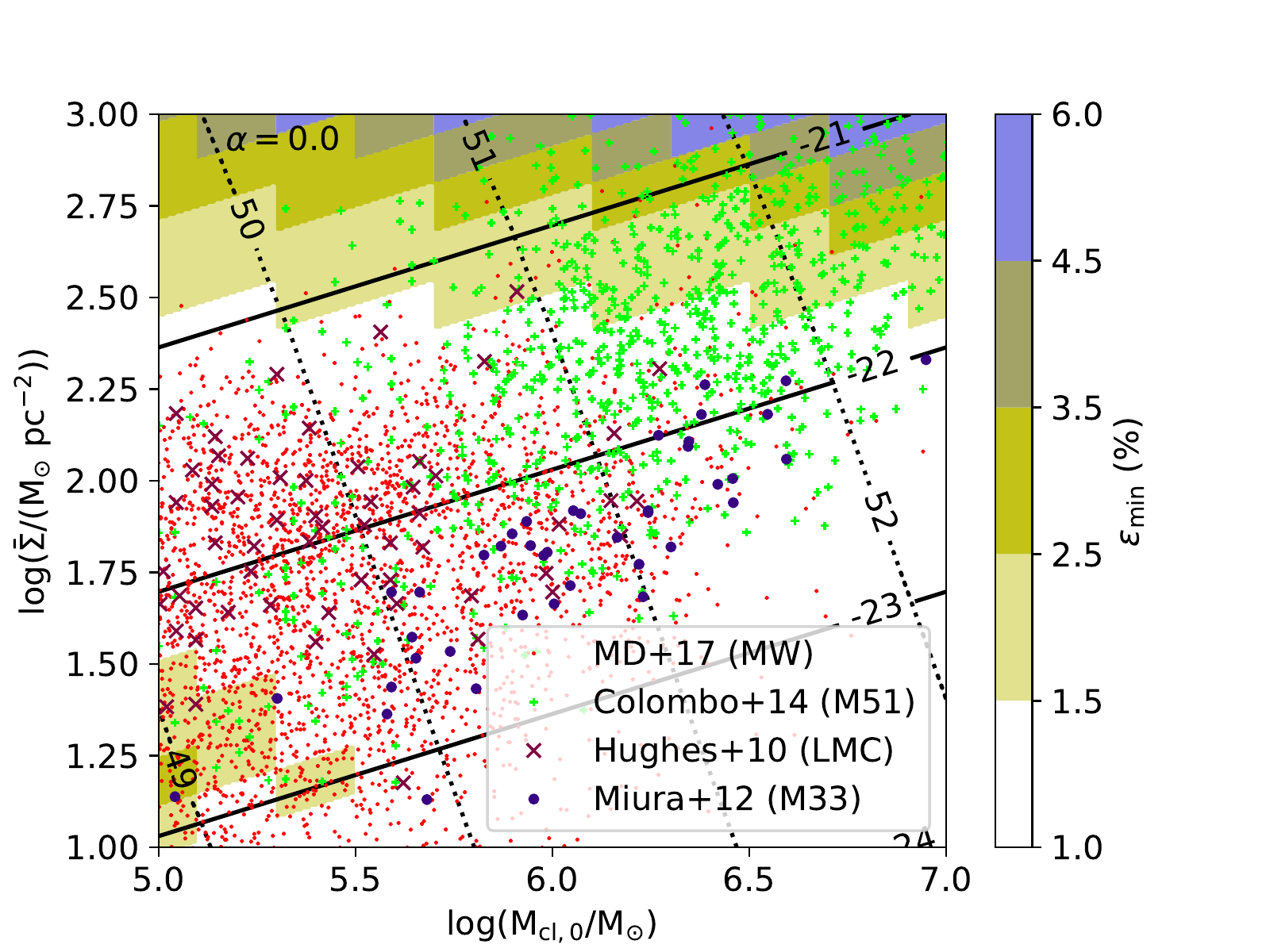}
\includegraphics[width=0.48\linewidth]{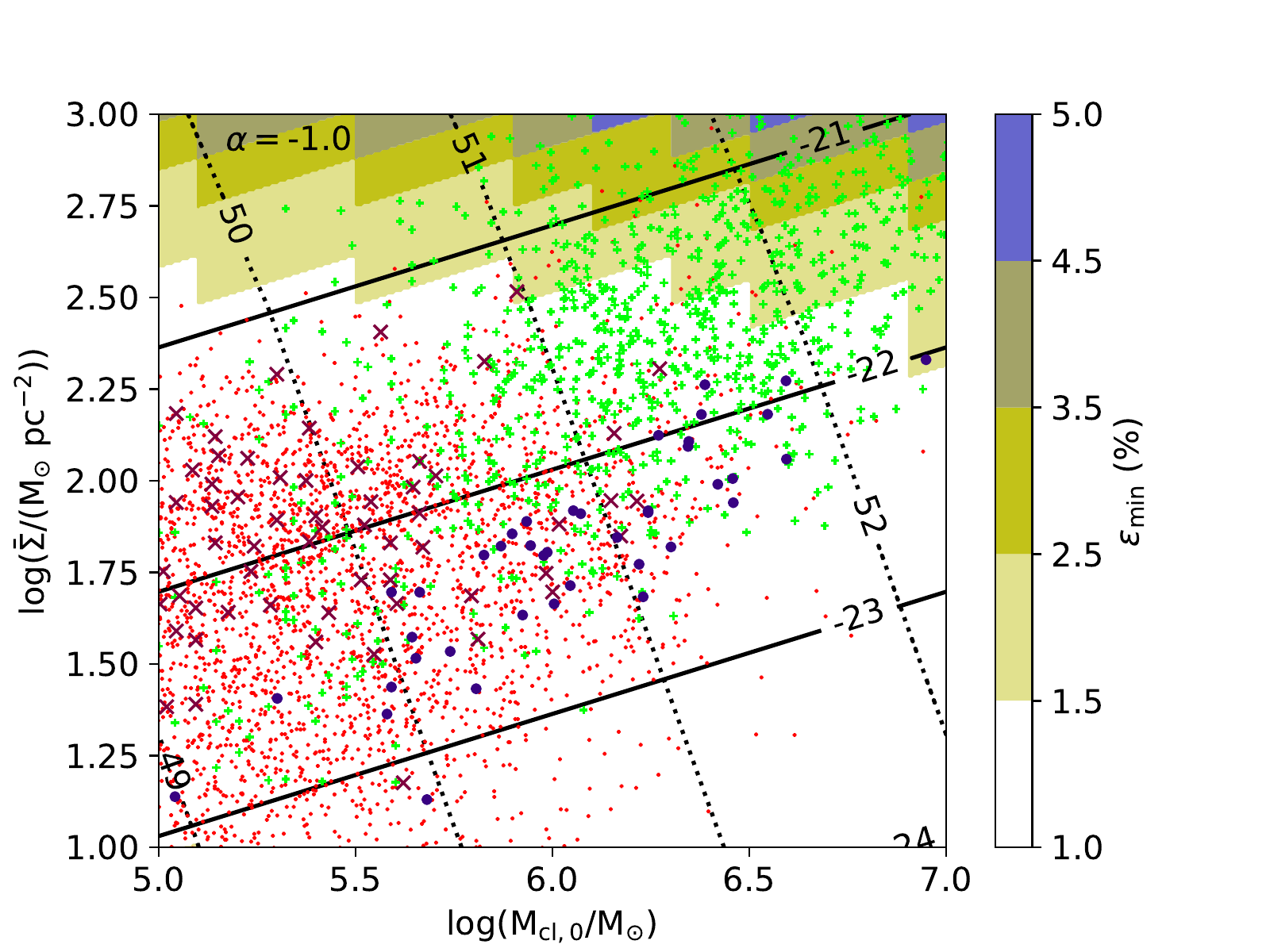}
\includegraphics[width=0.48\linewidth]{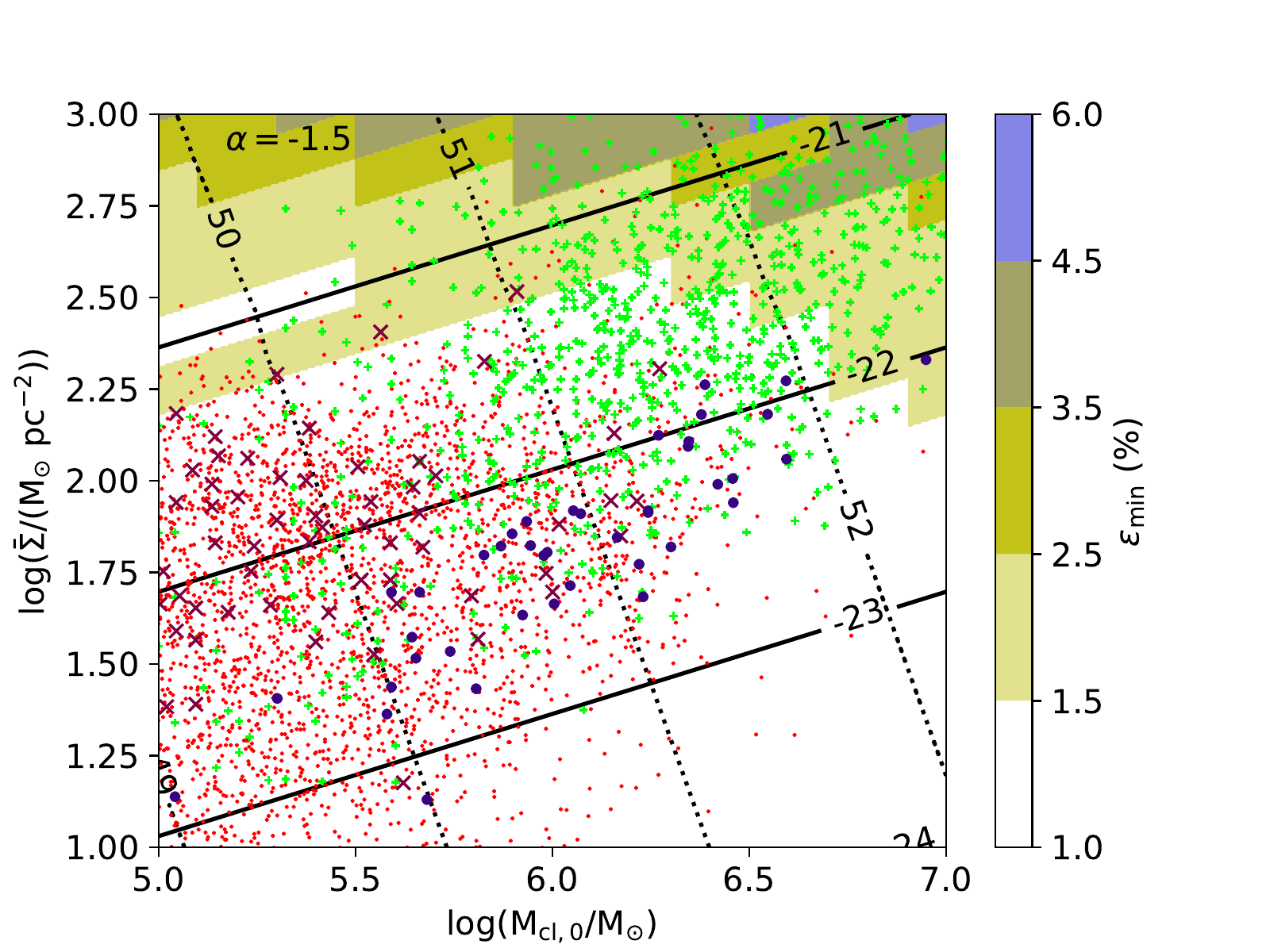}
\includegraphics[width=0.48\linewidth]{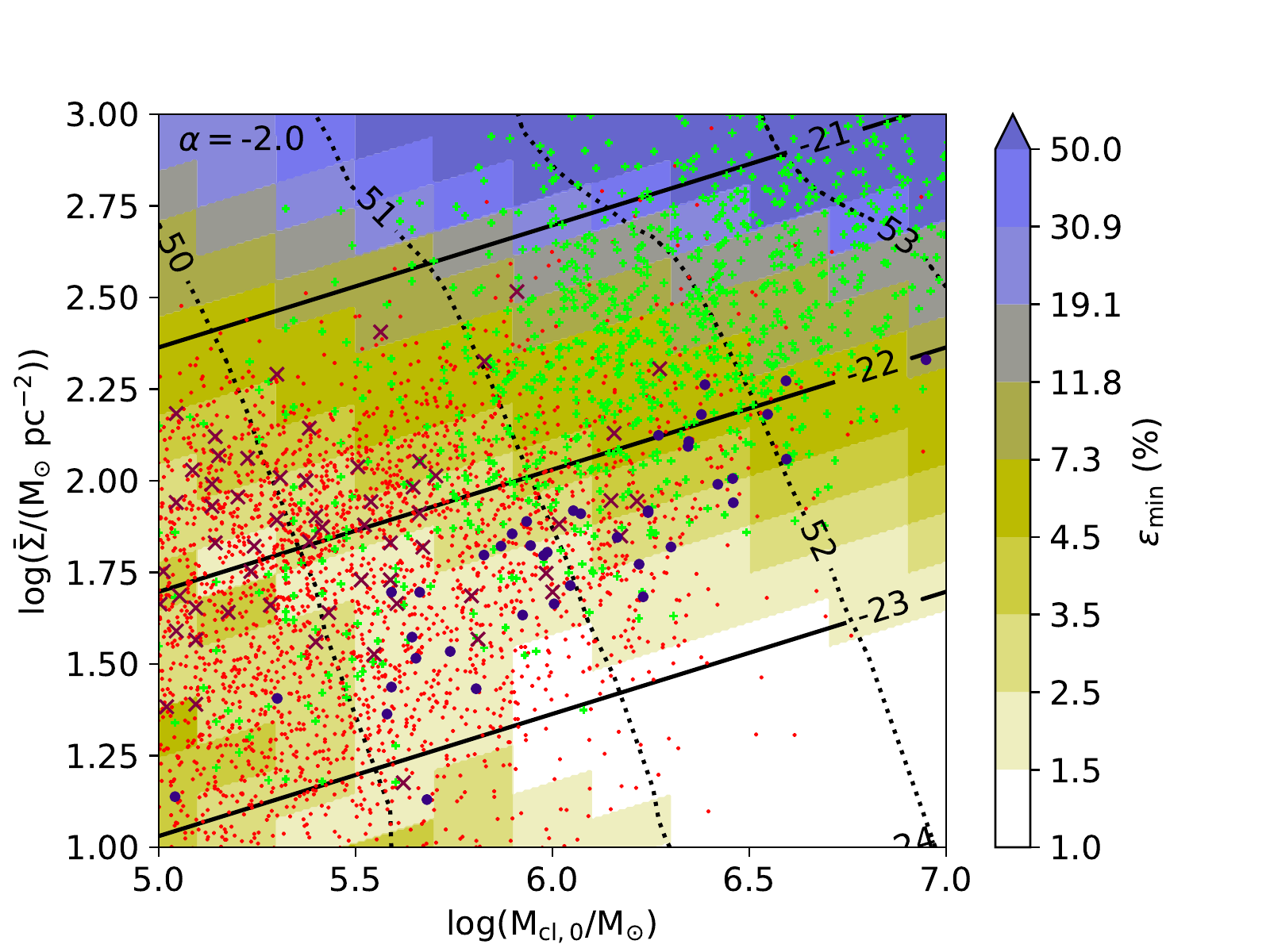}
\caption{Minimum star formation efficiencies $\SFEmin$ for different initial density slopes. Markers show observed GMCs in the Milky Way \citep{Miville-Deschenes2017}, M51 \citep{Colombo2014}, the LMC \citep{Hughes2010}, and M33 \citep{Miura2012}. Black solid lines show $\log \bar{\rho}$ in g\,cm$^{-3}$, black dotted lines show $\log(-E_{\mathrm{bind}})$ in erg. The sawtooth pattern is caused by our sampling of the parameter space (we use regular grids in $\log M_{\mathrm{cl},0}$ and $\log \bar{\rho}$ instead of $\log M_{\mathrm{cl},0}$ and $\log \bar{\Sigma}$). }
\label{fig:SFEmin}
\end{figure*}

Most GMCs in the Milky Way (MW) and other nearby galaxies have average surface densities in the range $10 \leq \bar{\Sigma} \leq 1000$\,M$_{\odot}$\,pc$^{-2}$ \citep{Heyer2009, Hughes2010, Miura2012, Hughes2013, Colombo2014, Miville-Deschenes2017}. If we assume that GMCs are gravitationally bound and that they continue forming stars until completely disrupted by stellar feedback, then we can use {\sc warpfield} to find the minimum star formation efficiencies of the GMCs as a function of their mass and surface density, i.e.\ the smallest possible fraction of their mass that they must convert into stars in order for them to be completely disrupted. In this section, we present results for clouds with surface densities in the observed range and cloud masses in the range $10^5 \leq M_{\mathrm{cl},0} \leq 10^7\,\Msun$.

We sample this regime by varying $M_{\mathrm{cl},0}$ and $\bar{\rho}$ with steps of $\Delta \left(\log M_{\mathrm{cl},0}\right) = \Delta \left(\log \bar{\rho}\right) = 0.2$. The average surface density is related to average mass densities via
\begin{equation} \label{avgdens_avgsurfdens}
\frac{4\pi}{3}\Rcl^3 \bar{\rho} = M_{\mathrm{cl}} = \pi \Rcl^2 \bar{\Sigma} .
\end{equation}
We also consider several different density slopes, $\alpha = 0, -1, -1.5$ and $-2$. For each set of input parameters we determine $\SFEmin$ to a precision of 1\,\%, rounded up.

The derived minimum star formation efficiencies for different density slopes are presented in Fig.~\ref{fig:SFEmin}. As shown, for fixed $\alpha$ it is good first approximation to treat $\SFEmin$ as independent of $\Mcl$, when assuming $\bar{\Sigma}$ does not, or only very weakly, depend on the cloud mass as implied by the third Larson relation \citep{Larson1981}. Instead the minimum star formation efficiency is mainly determined by $\bar{\Sigma}$, as already argued by \citet{Fall2010}. We note that the values for $\SFEmin$ calculated here are lower than those derived in \citet{Kim2016}, who ignore mechanical feedback, and \citet{Fall2010}, who regard only comparatively small contributions by winds, SNe and indirect radiation pressure to the total feedback budget as reasonable. The presented minimum star formation efficiencies are also lower than the values calculated in \citet{Rahner2017} using {\sc warpfield1} owing to the change in our treatment of cooling of the wind bubble and a sampling of the IMF up to 120\,$\Msun$.\footnote{In \citet{Rahner2017}, 100\,$\Msun$ was used as the upper mass cut-off of the ISM, i.e.\ feedback from the most massive stars was missing.} We also note, that if -- in contradiction to the third Larson relation -- more massive clouds also have a significantly higher mean surface density \citep[e.g.][see also Fig.~\ref{fig:SFEmin}]{Colombo2014, Miura2012}, we expect the star formation efficiency in massive GMCs to be higher than in low mass GMCs \citep[see also][]{Rahner2017}.

Even though in our model, star formation occurs in a short burst\footnote{In the case of re-collapse, we instead have several distinct bursts, but we do not discuss this case here.} and is not spread out over a long time period, we can approximate a star formation efficiency per free-fall time \citep[see][]{Krumholz2005} by using the lifetime $t_{\mathrm{life}}$ of the cloud as the relevant time scale, i.e.\
\begin{equation}
\SFEff \approx \SFE \frac{\tff}{t_{\mathrm{life}}} ,
\end{equation}
where $\tff$ is calculated from the average cloud density.
If we assume that clouds form stars with a total star formation efficiency of $\SFE = \SFEmin$, most of our models have $\SFEff \sim 0.3$\,\% in good agreement with observations on GMC scales \citep{Leroy2017}. 

 It is notable that with few exceptions clouds with $\bar{\Sigma} \leq 100\,\Msun$\,pc$^{-2}$ and $\alpha \geq -1.5$ can be destroyed if $\SFE \geq \SFEmin \approx 1$\,\%. Such a low value for the minimum star formation efficiency is a strong indicator that stellar feedback alone is sufficient to explain the low observed star formation efficiencies of GMCs -- which is not to say that in reality other physical processes like turbulence of the ISM \citep{Elmegreen2004,MacLow2004} and to a lesser degree magnetic fields \citep{Shu1987} do not also play a role in bringing $\SFE$ down.
 
 We note, however, that if clouds form stars preferentially in one short burst with $\SFE \approx \SFEmin$, this presents a challenge to the survivability of these star clusters because star clusters with $\SFE \lesssim 10$\,\% tend to dissolve after the natal gas has been removed \citep{Baumgardt2007, Shukirgaliyev2017}. We speculate that studies of star clusters with N-body simulations that include a centrally peaked star-formation profile as in the model by \citet{Parmentier2013} together with slow gas removal with velocities of approximately 10\,km\,s$^{-1}$ as predicted by our \textsc{warpfield2} simulations will result in a higher survival rate even for low star formation efficiencies.
 
\subsubsection{Giant Clumps}
 
Giant clumps at redshift $z\sim 1-3$ with gas masses of $10^7 - 10^9\,\Msun$ have typical surface densities of $\bar{\Sigma} \sim 100\,\Msun$\,pc$^{-2}$ and are consistent with isothermal spheres, i.e. $\alpha = -2$ \citep{Ceverino2012}. In this paper we probe the lower mass end of such clumps. Our results indicate that they are more resilient against destruction by stellar feedback than clouds with shallower density profiles.
As noted above, this is due to the larger binding energy of the giant clump and in particular of the inner region when $\alpha = -2$ (see Eq.~\ref{Ebind}). As more of the gas mass is accumulated in the inner regions, an expanding shell in a cloud with such a steep density profile decelerates faster, making it more susceptible to energy loss via leakage of hot gas after gravitational fragmentation and via strong cooling. 

 Our results indicate that giant clumps at the low mass end ($M_{\mathrm{cl},0} = 10^7\,\Msun$) can be destroyed by a single star burst with $\SFE \geq 4$\,\% for $\bar{\Sigma} = 100\,\Msun$\,pc$^{-2}$ ($\SFEmin = 1-12\,$\% for $30 \leq \bar{\Sigma} \leq 300\,\Msun$\,pc$^{-2}$). Were the star formation spread out over the lifetime of the clump ($\sim 30$\,Myr), this would correspond to $\SFEff = 0.5$\,\% ($0.3-0.7\,$\% for $30 \leq \bar{\Sigma} \leq 300\,\Msun$\,pc$^{-2}$). This result is in line with simulations by \citet{Oklopcic2017} but puts analytic models by \citet{Krumholz2010} into question who argue that star formation efficiencies per free-fall time of a few percent are insufficient to disrupt a giant molecular clump (they focus on clumps with  $M_{\mathrm{cl},0} \geq 3\times 10^7\,\Msun$ though). Our result does not yet take into account that the metallicity in galaxies at $z\sim 2$ is approximately a factor 2 lower than in present-day galaxies \citep{Yuan2012}, which will affect the amount of radiative cooling, the strength of metal-line driven winds of massive stars and the coupling of radiation to the ISM. We plan to revisit giant clumps with \textsc{warpfield} in the future.

\section{Conclusion} \label{sec:conclusion}

In this paper we have presented improvements to the 1D stellar feedback code \textsc{warpfield} \citep{Rahner2017}. \textsc{warpfield} is a fast, publicly available code which models the formation of a shell around a massive cluster and how this shell is affected by stellar winds, radiation, and SNe, as well as gravity. The improvements which are part of a new code release, \textsc{warpfield2}, include, but are not limited to, a better treatment of the early, energy-driven expansion phase of shells around massive clusters, their fragmentation, and the cooling of hot wind bubbles. 

In order to model the cooling in the wind bubble correctly we have produced a large grid of cooling curves with \textsc{cloudy} \citep{Ferland2017} that account for the ionizing radiation produced by the young massive cluster creating the wind bubble in additional to collisional ionization. The grid, which encompasses a wide range of temperatures, densities, photon fluxes of ionizing radiation, stellar ages, and two different metallicities (see Table~\ref{tab:cooling_grid} and Appendix \ref{sec:cooling_appendix}) is publicly available.

We have employed \textsc{warpfield2} to model the destruction of GMCs with various density profiles. Our main results are summarized as follows.

\begin{itemize}
\item With respect to the previous release, \textsc{warpfield1}, feedback of a young massive cluster is more efficient in destroying the parental molecular cloud.
\item Clouds with density profiles $\rho \propto R^{-\alpha}$ and a constant density core are investigated. We find that clouds with $-1.5 \leq \alpha \leq 0$ react very similarly to feedback if the average density of the cloud is kept constant.
\item The minimum star formation efficiency $\SFEmin$ needed to destroy a cloud after a single star burst is between $1 - 6$\,\% for GMCs with $-1.5 \leq \alpha \leq 0$ and with mean surface densities $\bar{\Sigma}$ of $10 - 1000\,\Msun$\,pc$^{-2}$. The value of $\SFEmin$ is mainly set by $\bar{\Sigma}$. Varying the cloud mass is a second order effect.
\item Typical star formation efficiencies per free-fall time are $\sim 0.3$\,\%, in good agreement with observations on GMC scales by \citet{Leroy2017}.
\item For $\alpha = -2$ (as suggested for giant clumps at $z\sim 2$) $\SFEmin$ can be much higher (from $1\,$\% for $\bar{\Sigma} \sim 10\,\Msun$\,pc$^{-2}$ to $\gtrsim 50\,$\,\% for $\bar{\Sigma} \sim 1000\,\Msun$\,pc$^{-2}$). At a fiducial surface density of $100\,\Msun$\,pc$^{-2}$ we predict $\SFEmin \sim 5\,$\%.
\item Stellar feedback alone is sufficient in explaining the low observed star formation efficiencies in star forming regions.
\end{itemize}

\section*{Acknowledgements}

We acknowledge funding from the Deutsche Forschungsgemeinschaft in the Collaborative Research Centre (SFB 881) ``The Milky Way System'' (subprojects B1, B2, and B8), the Priority Program SPP 1573 ``Physics of the Interstellar Medium'' (grant numbers KL 1358/18.1, KL 1358/19.2, and GL 668/2-1), and the European Research Council in the ERC Advanced Grant STARLIGHT (project no. 339177). This research made use of \textsc{scipy} \citep{jones_scipy_2001} and \textsc{matplotlib} \citep{Hunter2007}.

\begin{footnotesize}
\bibliographystyle{mn2e}  
\bibliography{libfinal}
\end{footnotesize}

\appendix

\section{Bubble Structure} \label{sec:Bubble_Struc}
When thermal conduction and radiative cooling are included in the self-similarity solution of \citet{Weaver1977}, the velocity structure and the temperature structure of the bubble are given by the set of differential equations, 
\begin{eqnarray} \label{structure_vel}
v' &=& \frac{\beta + \delta}{t} + \left(v-\frac{\alp r}{t} \right) \frac{T'}{T} - \frac{2v}{r},  \\ \nonumber
T'' &=& \frac{P_{\rm b}}{C T^{5/2}} \left[ \frac{\beta + 2.5\delta}{t} + \frac{P_{\rm b}}{4k^2}\frac{\Lambda_{\mathrm{net}}}{T^2} + 2.5\left( v-\frac{\alp r}{t}\right)\frac{T'}{T}\right]  \\ \label{structure_temp}
&& - \frac{2.5T'^2}{T}-\frac{2T'}{r},
\end{eqnarray}
where $v$ is the gas velocity and where primes indicate differentiation with respect to the radius $r$. Here,
\begin{eqnarray}
\alp &\equiv & \frac{\partial \ln R_2}{\partial \ln t}, \\ \label{beta}
\beta &\equiv & -\frac{\partial \ln P_{\rm b}}{\partial \ln t}, \\
\delta &\equiv & \left(\frac{\partial \ln T}{\partial \ln t}\right)_\xi,
\end{eqnarray}
with $\xi = r/R_2$, and $k$ is the Boltzmann constant. We use a constant thermal conduction coefficient of $C = 6\times 10^{-7}$ erg\,s$^{-1}$\,cm$^{-1}$\,K$^{-7/2}$. It is related to the thermal conductivity $\kappa$ via
\begin{equation}
\kappa = C\cdot T_{\mathrm{e}}^{5/2},
\end{equation}
where $T_{\mathrm{e}}$ is the electron temperature. In reality, $C$ is not a constant but
\begin{equation} \label{conduction}
C = 4.6\times 10^{13}\left(10^8\,\textup{K}\right)^{-5/2} \left(\frac{\ln \Lambda}{40}\right)^{-1} \,\mathrm{s}^{-1}\,\mathrm{cm}^{-1}\,\mathrm{K}^{-7/2},
\end{equation}
where the Coulomb logarithm is \citep{Cowie1977}
\begin{equation}
\ln \Lambda = 29.7 + \ln \left( \sqrt{\frac{1\,\mathrm{cm}^{-3}}{n}} \frac{T_{\mathrm{e}}}{10^6\,\mathrm{K}}\right) \quad \textup{for}\ T > 4.2 \times 10^5\,\mathrm{K}.
\end{equation} 
For reasonable values of $n$ and $T_{\mathrm{e}}$, $C$ as derived from eq. (\ref{conduction}) does not differ by more than 30\,\% from the constant value cited above \citep{Spitzer1956}.

The boundary conditions (BCs) for solving the structure equations (\ref{structure_vel}) and (\ref{structure_temp}) are 
\begin{eqnarray}
\lim\limits_{r\rightarrow R_2} T &=& 0, \\
\lim\limits_{r\rightarrow R_2} v &=& \dot{R}_2, \\
\lim\limits_{r\rightarrow R_1} v &=& 0.
\end{eqnarray}
Eqs. (\ref{structure_vel}) and (\ref{structure_temp}) are solved from $R_2$ to $R_1$ via a shooting method, that is, a guess is made for the remaining BC at $R_2$. If the solution of the ODE system then also fulfils the BCs at $R_1$, the choice of the right-hand side BC was correct. Otherwise, another guess is made and the correct right-hand side BC is determined via a root finding algorithm. More details are provided in \citet{Weaver1977}.

In order to determine $\beta$, it is necessary to know the value of $\dot{P}_{\rm b}$ (see Eq.~\ref{beta}), but the energy equation (Eq.~\ref{energy_w_cooling}) only gives an expression for $\dot{E}_{\rm b}$. To close the system of equations a relation between $\dot{P}_{\rm b}$ and $\dot{E}_{\rm b}$ must be derived. This relation follows from Eqs.~(\ref{pressure_correct}) and (\ref{R1}):
\begin{equation}
\dot{E}_{\rm b} = \frac{2\pi \dot{P}_{\rm b} d^2 + 3E_{\rm b}\dot{R}_2 R_2^2 \left( 1 - \frac{c}{E_{\rm b}+c}\right) - a\frac{R_1^3 E_{\rm b}^2}{(E_{\rm b}+c)}}{d\left( 1- \frac{c}{E_{\rm b}+c}\right)}
\end{equation}
with 
\begin{eqnarray}
a &\equiv& \frac{3}{2}\frac{\dot{F}_{\mathrm{ram}}}{F_{\mathrm{ram}}}, \\
c &\equiv& \frac{3}{4} F_{\mathrm{ram}} R_1, \\
d &\equiv& R_2^3 - R_1^3.
\end{eqnarray}
In order to solve the ODE system governing the dynamics of the bubble, Eqs. (\ref{momentum_full}), (\ref{energy_w_cooling}), and (\ref{Txi}), for the first time step we start with the self-similarity values for $\alp, \beta$, and $\delta $ (where cooling is not included), i.e. $\alp = 3/5$, $\beta = 4/5$, $\delta = -6/35$ \citep{Weaver1977}.

\section{Cooling curves} \label{sec:cooling_appendix}
For this paper 144,408 \textsc{cloudy} models have been run to derive cooling and heating values for a wide parameter range (see Table~\ref{tab:cooling_grid}). As an example, we show in Table~\ref{tab:cooling_grid_results} the heating and cooling rates per unit volume for a selection of different densities, temperatures and ionizing photon fluxes for the case of a 
1\,Myr-old star cluster with solar metallicity. The net cooling rate utilized in {\sc warpfield2} is simply the cooling rate minus the heating rate and hence is not tabulated. For an arbitrary combination of $t_{\mathrm{age}}$, $n$, $T$, and $\Phi_{\mathrm{i}}$, cooling and heating rates can be derived from the tabulated values using quadrilinear interpolation. 
For $t_{\mathrm{age}} < 1\,$\,Myr we consider the same spectral shape of the star cluster as for $t_{\mathrm{age}} = 1\,$\,Myr. This constitutes only a minor error as the change of the spectrum at early times is small.  

\begin{table*}
\begin{tabular}{|c|c|c|c|c|c|c|}
\hline 
ID & $n$ & $T$ & $\Phi_{\mathrm{i}}$ & $n_{\mathrm{e}}$ & $n n_{\mathrm{e}} \Gamma $ & $n n_{\mathrm{e}} \Lambda $ \\ 

-- & (cm$^{-3}$) & (K) & (cm$^{-2}$\,s$^{-1}$) & (cm$^{-3}$) & (erg\,cm$^{-3}$\,s$^{-1}$) & (erg\,cm$^{-3}$\,s$^{-1}$) \\ 
\hline 
1   &    $1.0\times 10^{-4}$  &   $3.162\times 10^{3}$  &   $1.0\times 10^{0}$       & $4.4\times 10^{-5}$   &   $1.467\times 10^{-30}$   &   $3.737\times 10^{-33}$   \\
2   &    $3.162\times 10^{-4}$   &   $3.162\times 10^{3}$   &   $1.0\times 10^{0}$       & $1.01\times 10^{-4}$   &   $4.236\times 10^{-30}$   &   $2.628\times 10^{-32}$ \\
\ldots & \ldots & \ldots & \ldots & \ldots & \ldots & \ldots \\
33   &    $1.0\times 10^{12}$  &   $3.162\times 10^{3}$  &   $1.0\times 10^{0}$   &    $1.5\times 10^{5}$  &   $1.978\times 10^{-15}$  &   $5.039\times 10^{-8}$ \\
34   &   $1.0\times 10^{-4}$  &   $3.162\times 10^{3}$  &   $1.0\times 10^{1}$   & $5.29\times 10^{-5}$  &   $1.566\times 10^{-30}$  &   $4.766\times 10^{-33}$ \\
35   &   $3.162\times 10^{-4}$  &   $3.162\times 10^{3}$  &   $1.0\times 10^{1}$      & $1.19\times 10^{-4}$  &   $4.458\times 10^{-30}$  &   $3.242\times 10^{-32}$ \\
\ldots & \ldots & \ldots & \ldots & \ldots & \ldots & \ldots \\
12034 &  $1.0\times 10^{12}$  &   $3.162\times 10^{5}$  &   $1.0\times 10^{21}$      & $1.19\times 10^{12}$   &  $2.15\times 10^{-2}$  &   $1.697\times 10^{2}$
\end{tabular} 
\caption{Shortened cooling table for a star cluster with $t_{\mathrm{age}} = 1$\,Myr and $Z = 0.014$ (solar). ID: identification number of \textsc{cloudy} model, $n$: ion number density, $T$: temperature, $\Phi_{\mathrm{i}}$: number flux of ionizing photons, $n_{\mathrm{e}}$: electron number density, $n n_{\mathrm{e}} \Gamma$: change rate of internal energy density due to heating, $n n_{\mathrm{e}} \Lambda$: change rate of internal energy density due to cooling (see Eq. \ref{radenerg}). The three rightmost columns contain derived quantities. For the shape of the spectrum we consider a fully sampled Kroupa IMF. For a full table refer to the online data.
\label{tab:cooling_grid_results}}
\end{table*}

\begin{figure*}
\centering
\includegraphics[width=0.48\linewidth]{plots/cool/{cool_n0.0_R1.0}.pdf}
\includegraphics[width=0.48\linewidth]{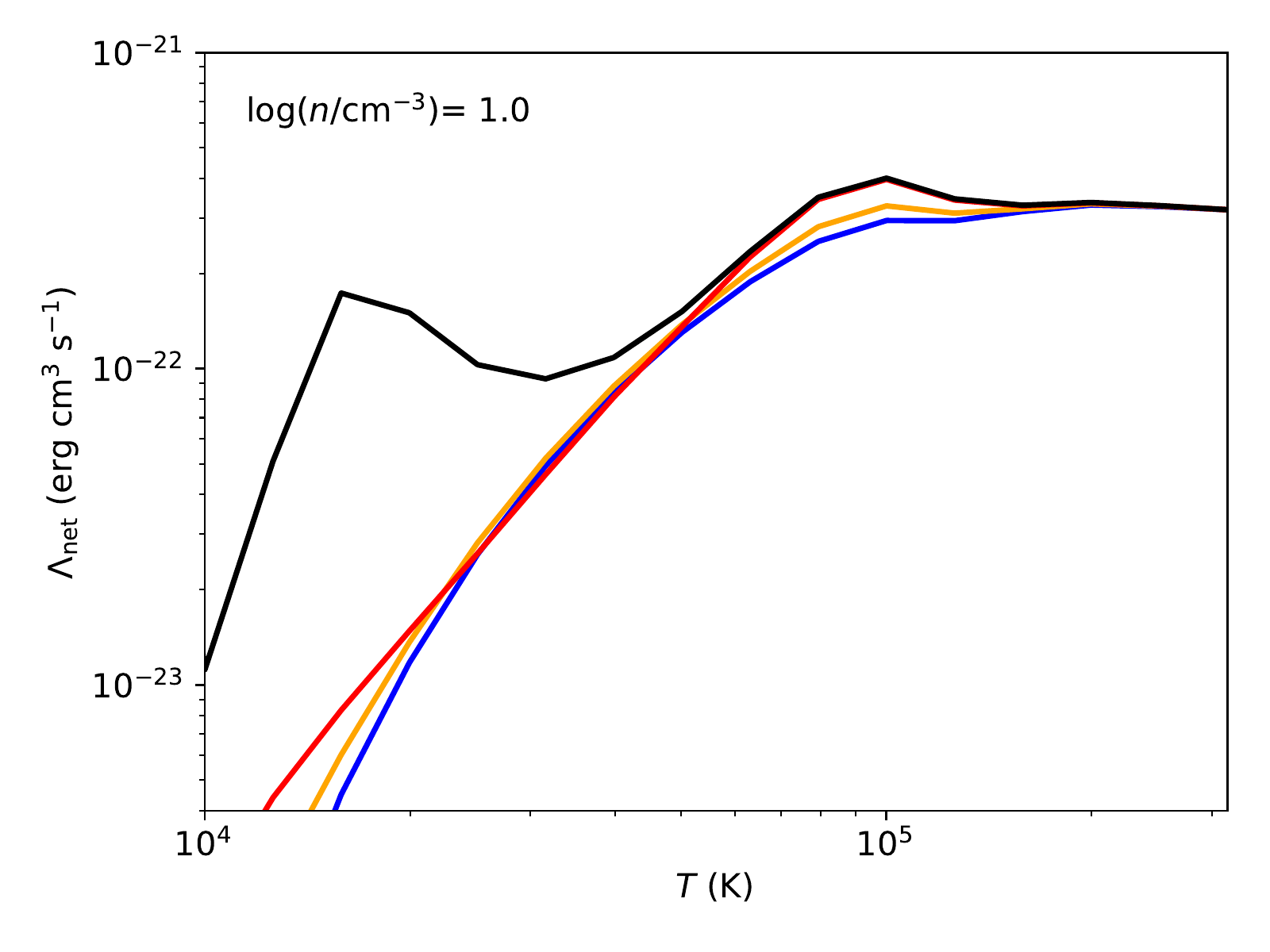}
\includegraphics[width=0.48\linewidth]{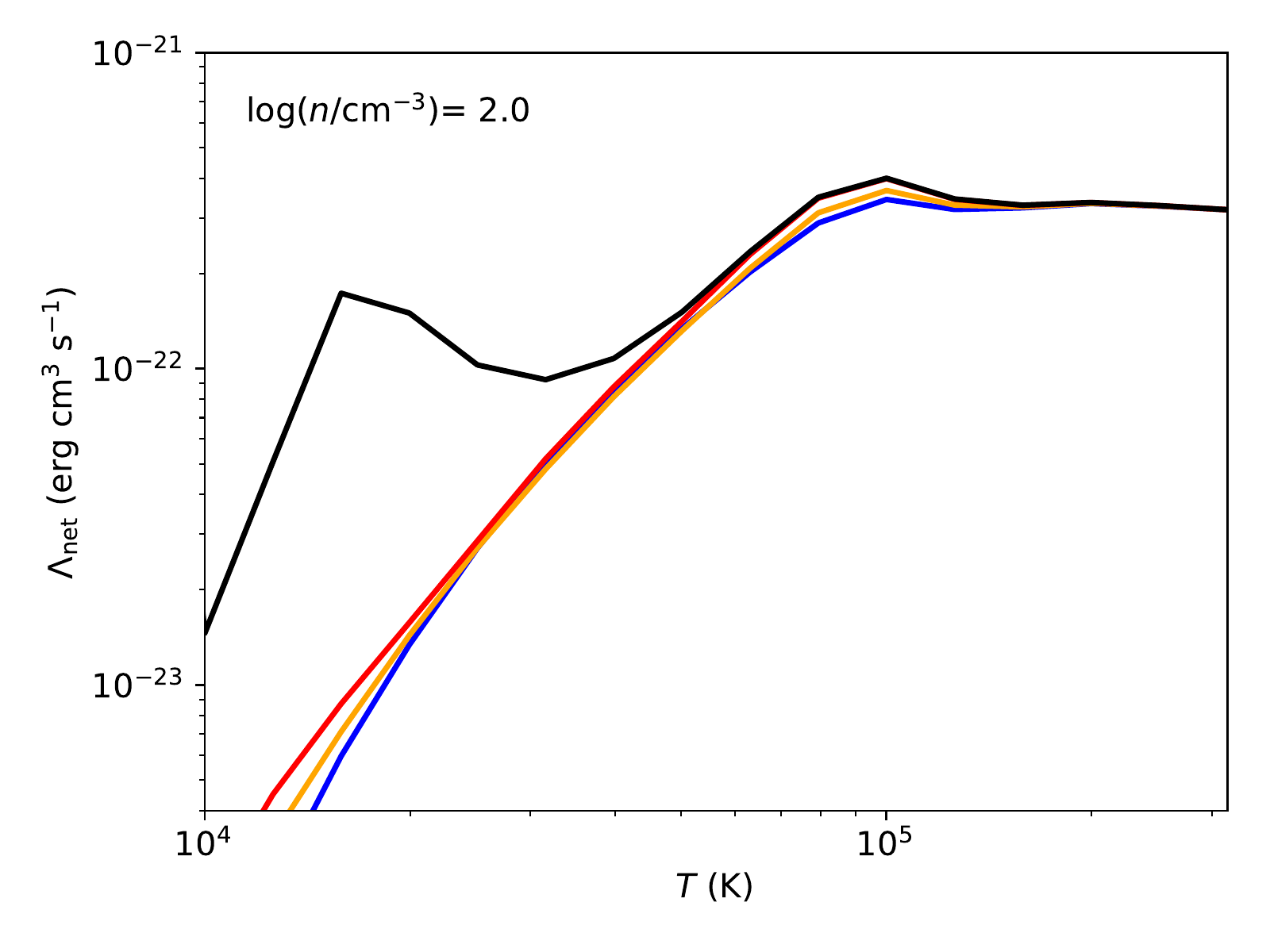}
\includegraphics[width=0.48\linewidth]{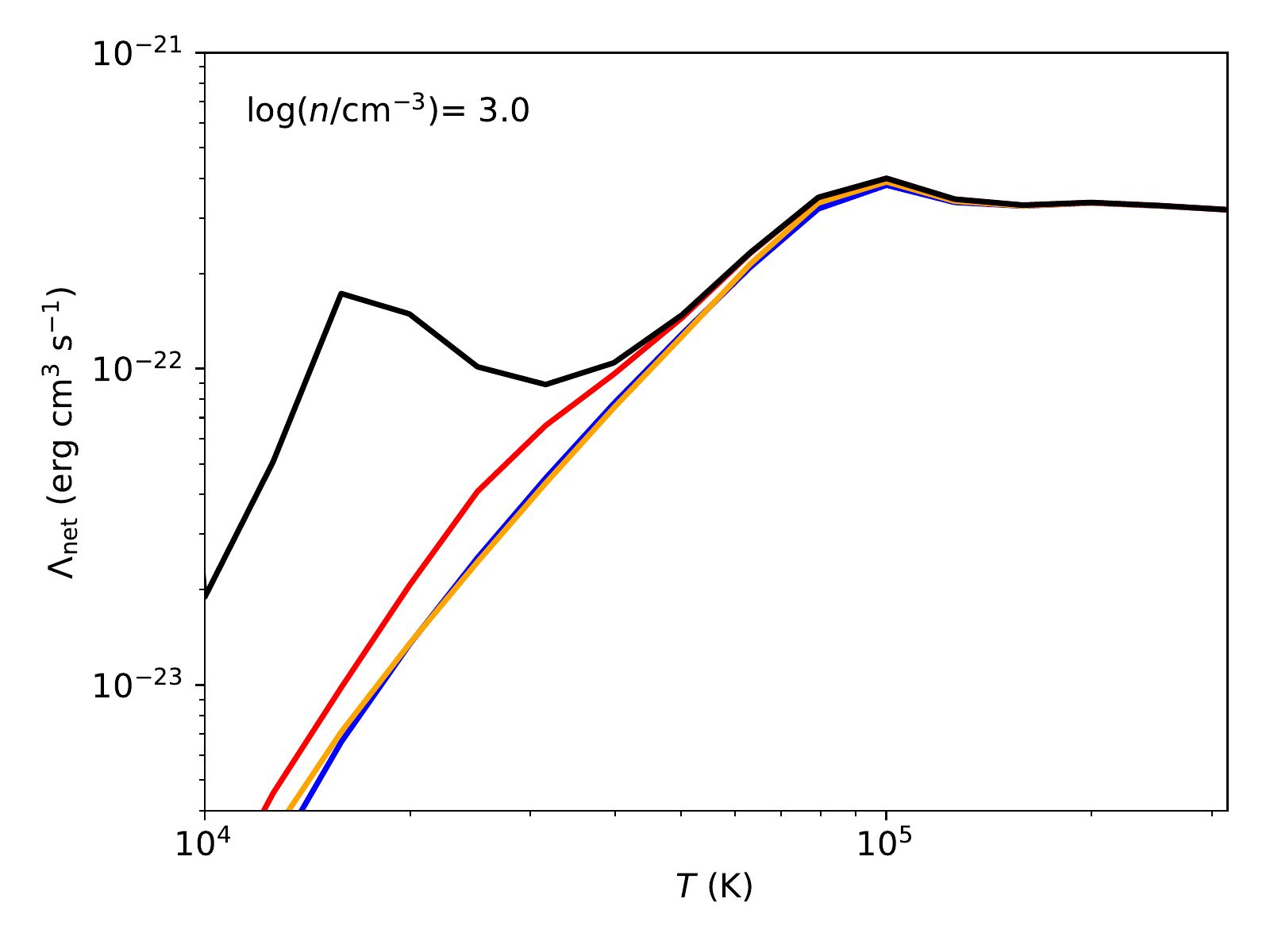}
\includegraphics[width=0.48\linewidth]{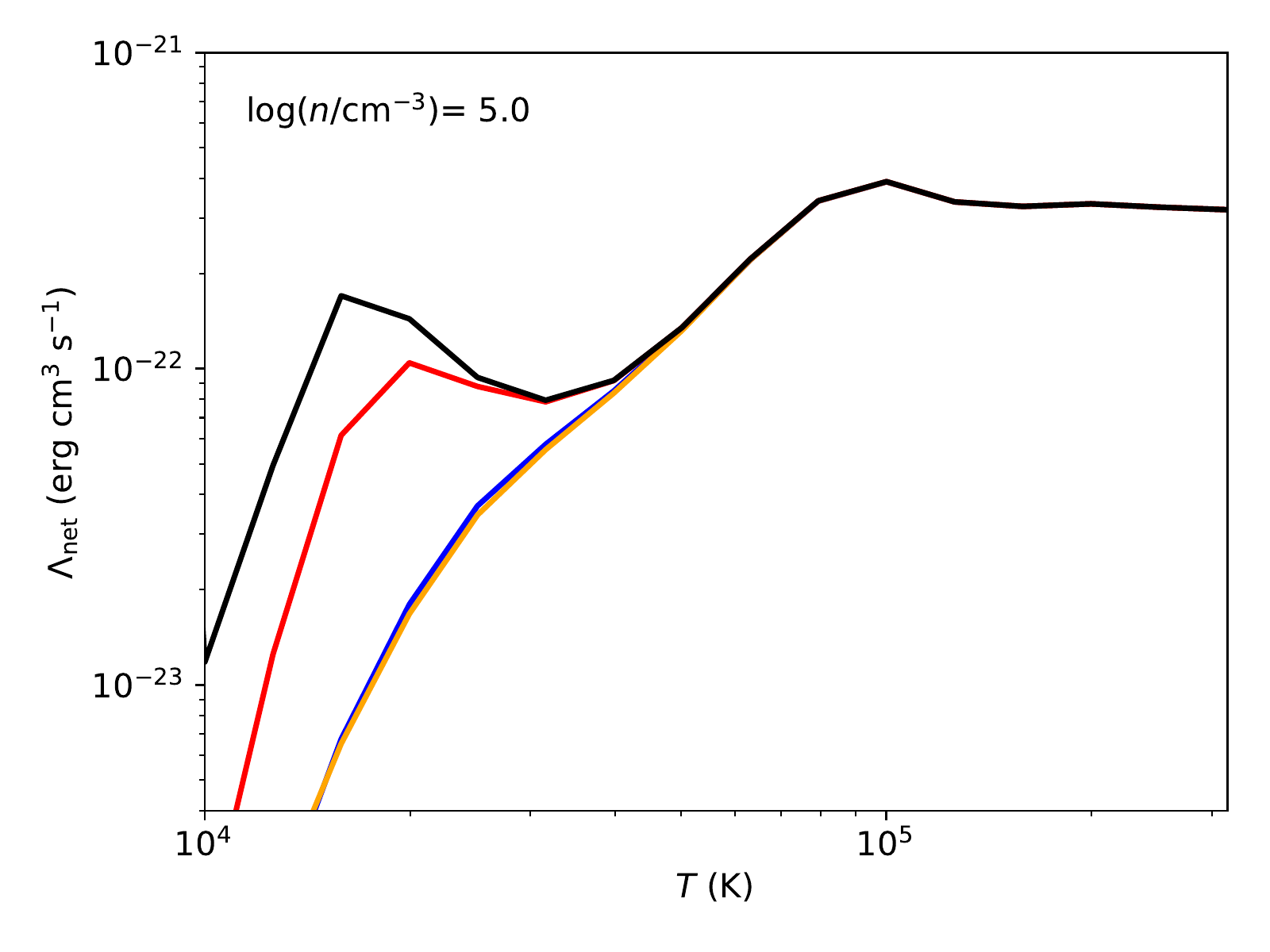}
\includegraphics[width=0.48\linewidth]{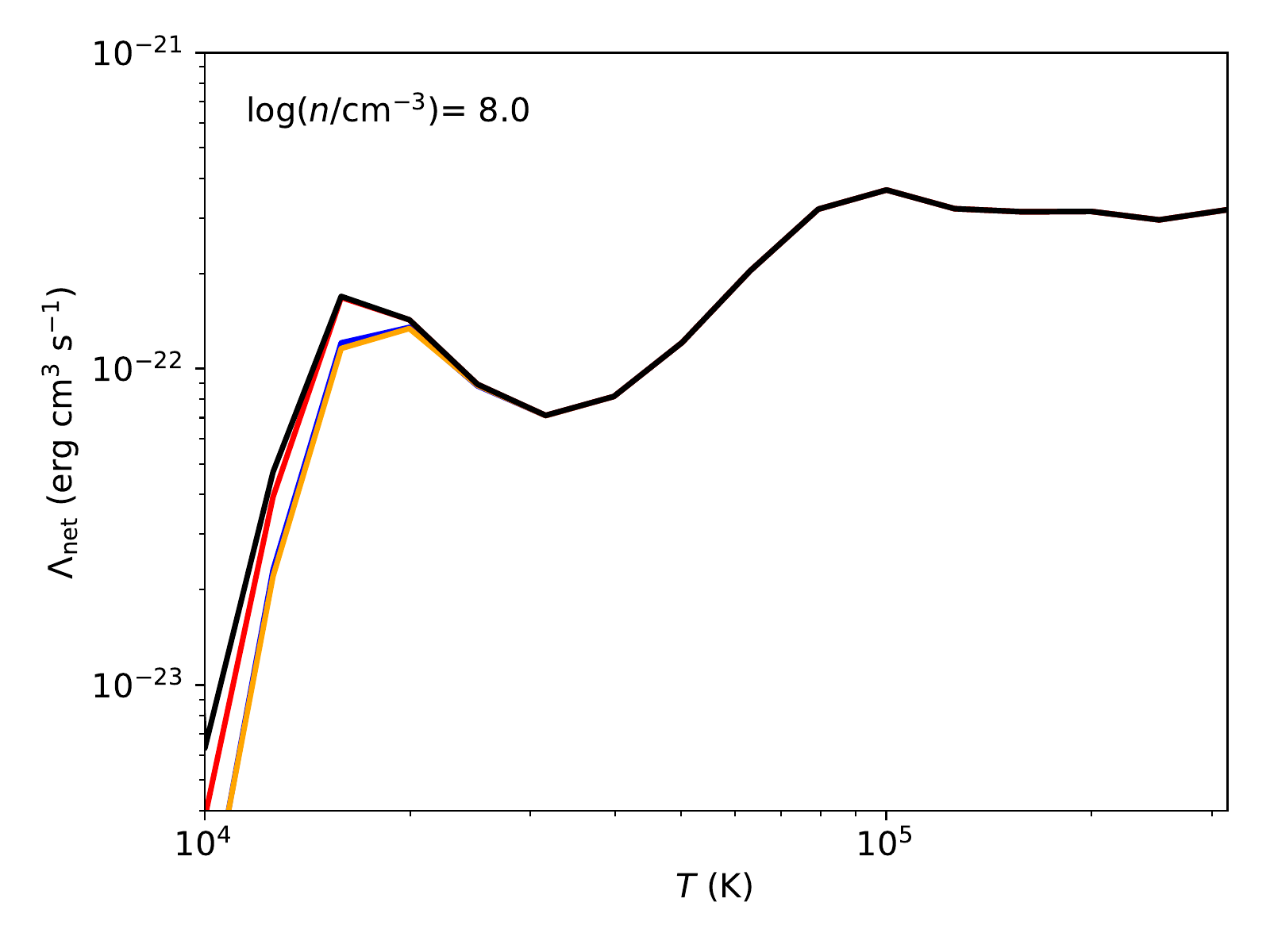}
\caption{Net cooling curves for solar metallicity (with elemental composition appropriate for HII regions) and various particle densities. Cooling curves are shown by different colours for different ages $t_{\mathrm{age}}$ of the illuminating star cluster ($M_* = 10^6$\,$M_{\odot}$). The ionizing photon flux ($\Phi_\mathrm{i,cgs} = \Phi_\mathrm{i}$/(cm$^{-2}$\,s$^{-1}$)) has been calculated according to the time evolution of the cluster and at a distance of 10\,pc. }
\label{fig:extra_cool10pc}
\end{figure*}

\begin{figure*}
\centering
\includegraphics[width=0.48\linewidth]{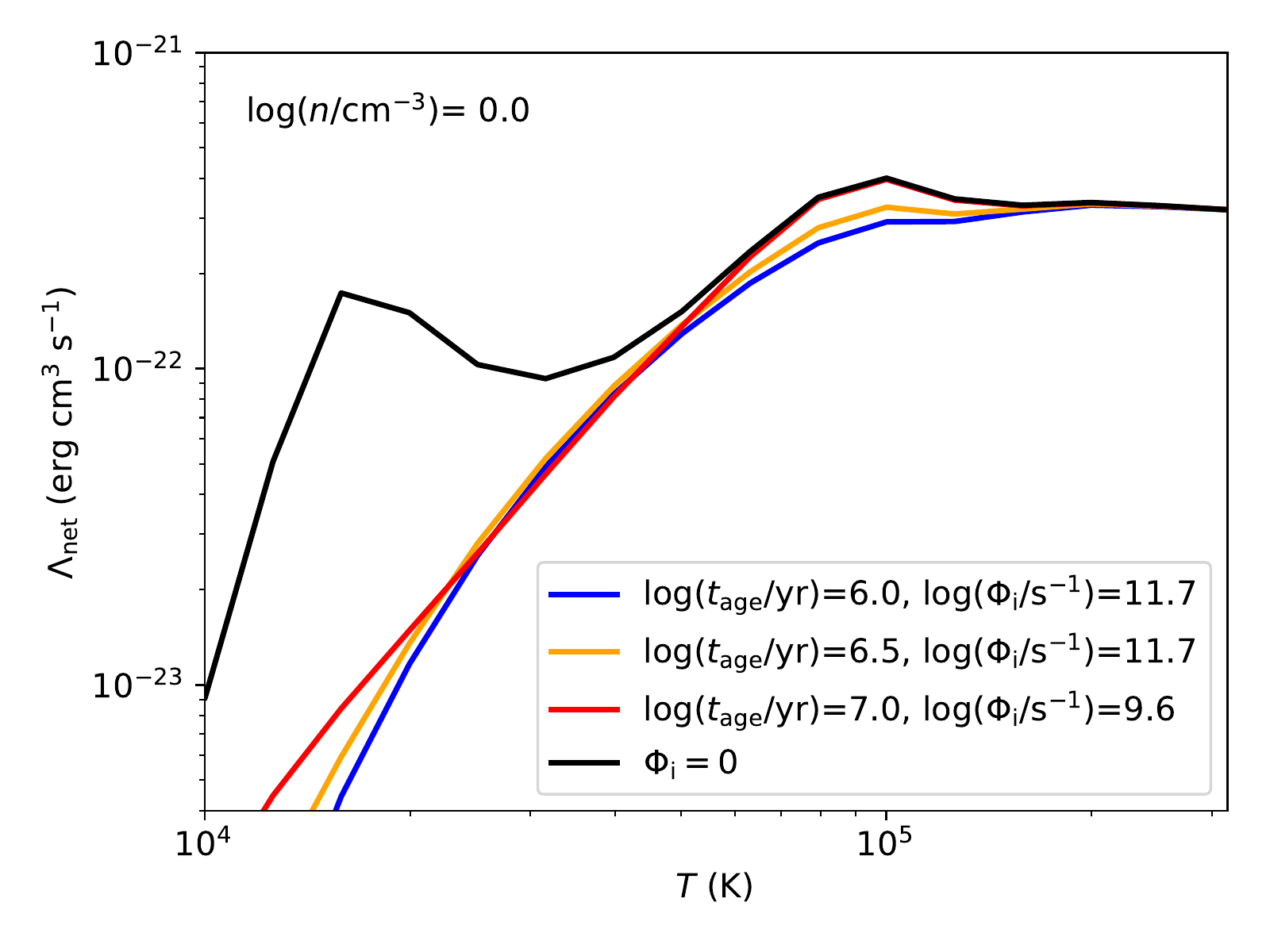}
\includegraphics[width=0.48\linewidth]{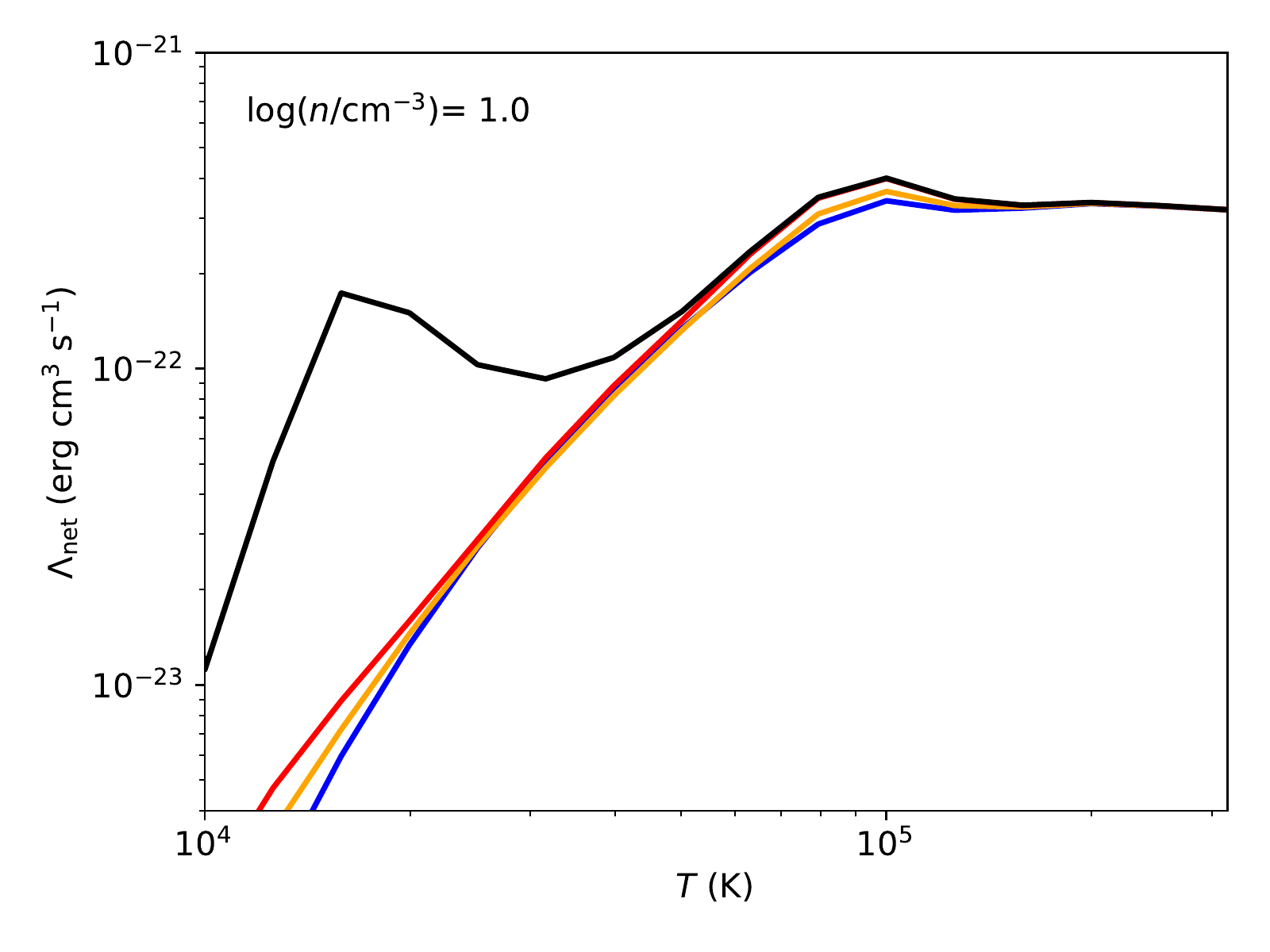}
\includegraphics[width=0.48\linewidth]{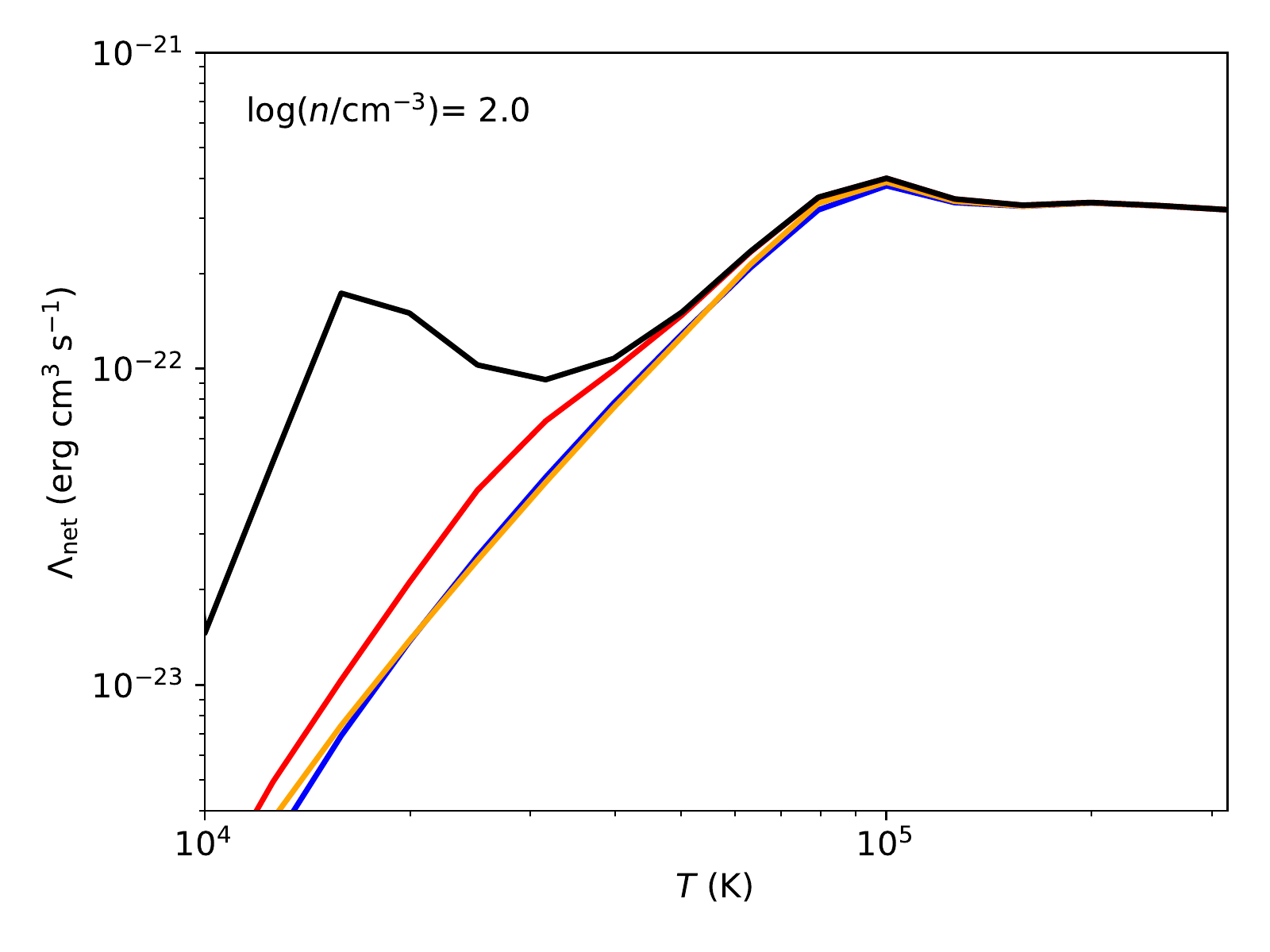}
\includegraphics[width=0.48\linewidth]{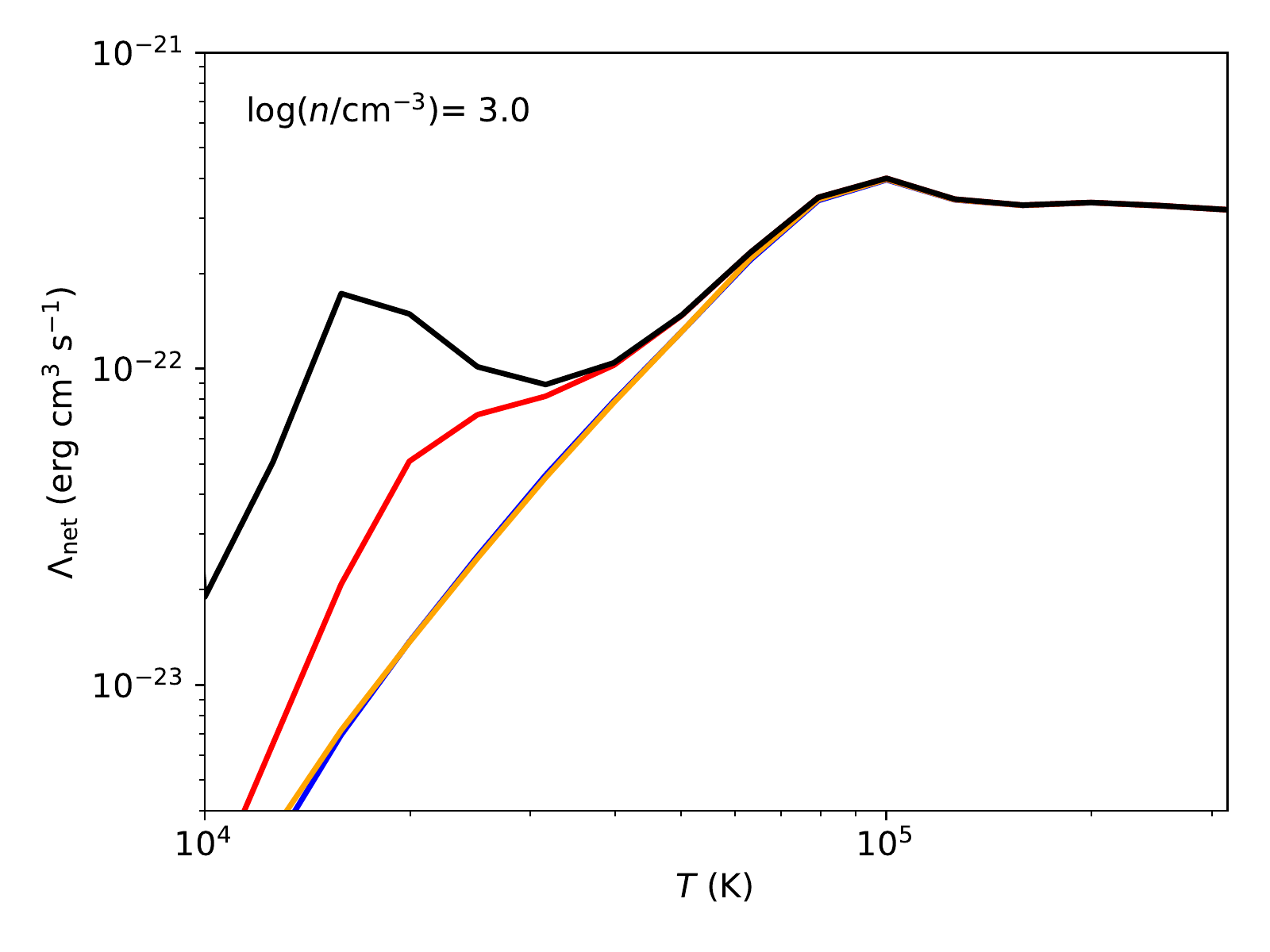}
\includegraphics[width=0.48\linewidth]{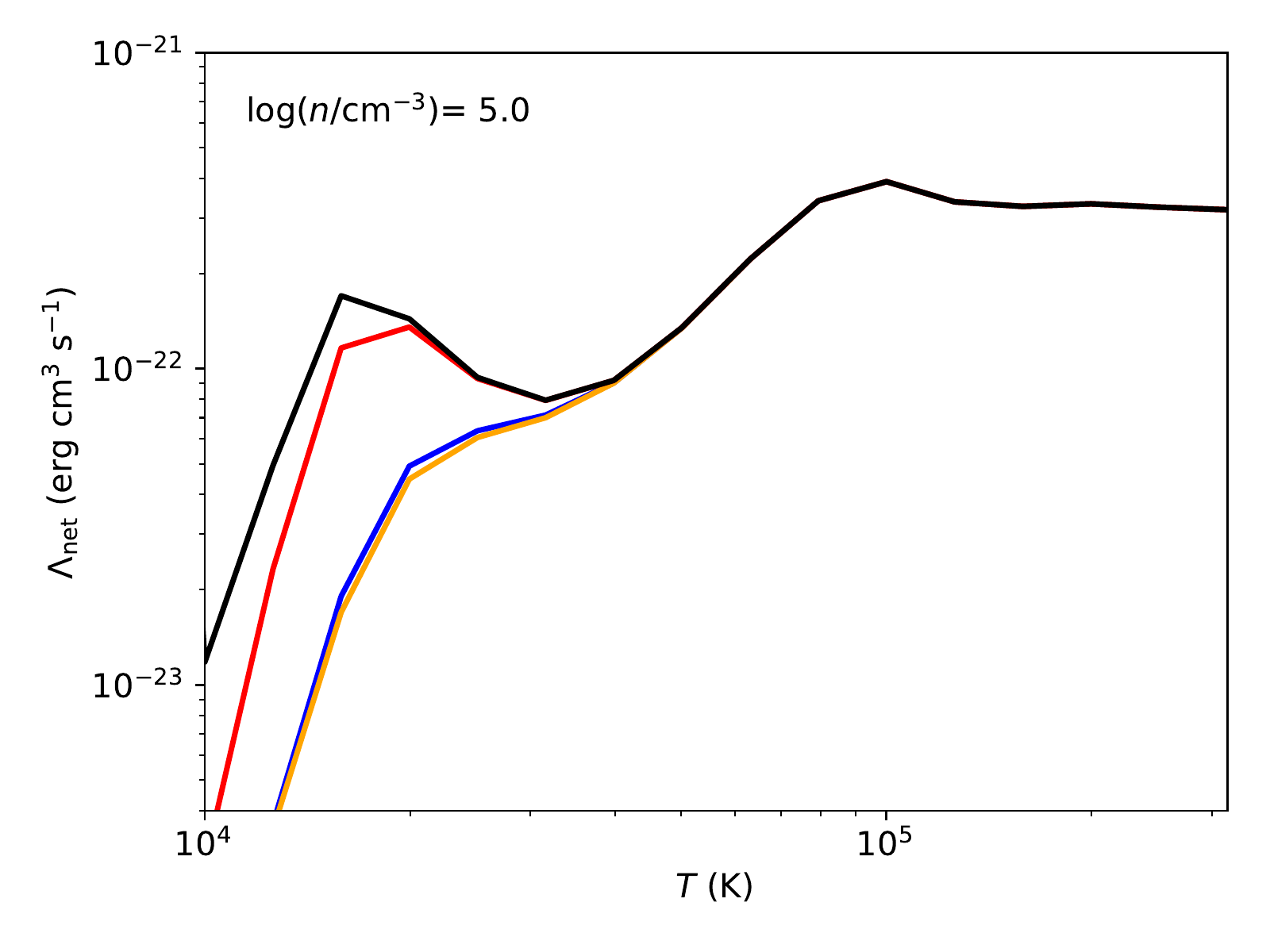}
\includegraphics[width=0.48\linewidth]{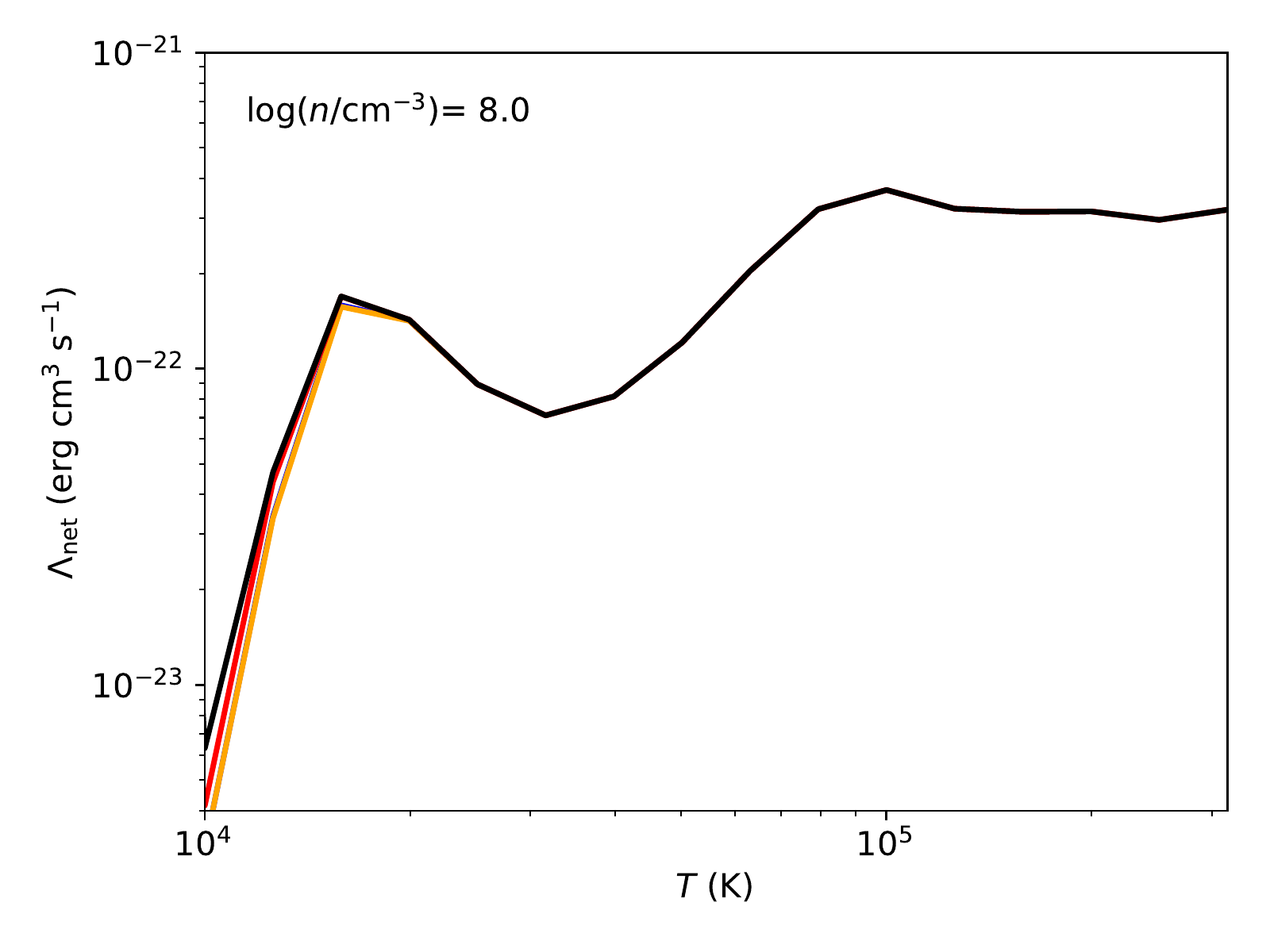}
\caption{Same as Fig. \ref{fig:extra_cool10pc} but here the distance between the star cluster and the illuminated ISM is 30\,pc.}
\label{fig:extra_cool30pc}
\end{figure*}

In Fig. \ref{fig:extra_cool10pc} and \ref{fig:extra_cool30pc} we present example cooling curves (showing the net cooling rate, $\Lambda_{\rm net}$) for the ISM at various densities and at distances of 10 and 30\,pc from an ageing star cluster ($t_{\mathrm{age}} = 1, 3, 10$\,Myr) with $M_* = 10^6\,\Msun$. For an older star cluster, the cooling curve approaches the CIE curve. Cosmic rays from the Galactic background are included, providing an extra source of heating and ionization, although they are a minor effect in the presence of a young massive cluster emitting a large number of ionizing photons. With growing age and decreasing mass of the cluster and with increasing distance from it, cosmic rays rise in importance.

For temperatures above $10^{5.5}$\,K, we use CIE cooling curves from \citet{Gnat2012} and \citet{Sutherland1993}.
The full data (including CIE curves at $T>10^{5.5}$\,K) are available online. 

\bsp    
\label{lastpage}
\end{document}